\documentclass[conference]{IEEEtran}
\IEEEoverridecommandlockouts
\usepackage{cite}
\usepackage{amsmath,amssymb,amsfonts}
\allowdisplaybreaks[4]
\usepackage[ruled]{algorithm2e}
\usepackage{multirow}
\usepackage{booktabs}
\usepackage{textcomp}
\usepackage{xcolor}
\usepackage{subcaption}
\usepackage{graphicx}
\usepackage{makecell}
\usepackage{verbatim} 
\usepackage{url}
\usepackage{threeparttable}

\def\BibTeX{{\rm B\kern-.05em{\sc i\kern-.025em b}\kern-.08em
    T\kern-.1667em\lower.7ex\hbox{E}\kern-.125emX}}
\begin{document}

\title{Tangram: High-resolution Video Analytics on Serverless Platform with SLO-aware Batching
}

\author{
	\IEEEauthorblockN{
		Haosong Peng\IEEEauthorrefmark{1}, 
		Yufeng Zhan\IEEEauthorrefmark{1}, 
		Peng Li\IEEEauthorrefmark{2}, 
		Yuanqing Xia\IEEEauthorrefmark{1} 
            } 
	\IEEEauthorblockA{\IEEEauthorrefmark{1}School of Automation, Beijing Institute of Technology, Beijing, China}
	\IEEEauthorblockA{\IEEEauthorrefmark{2}School of Computer Science and Engineering, The University of Aizu, Aizuwakamatsu, Japan}
 \IEEEauthorblockA{livion@bit.edu.cn, yu-feng.zhan@bit.edu.cn, pengli@u-aizu.ac.jp, xia\_yuanqing@bit.edu.cn}
} 

\maketitle

\begin{abstract}
Cloud-edge collaborative computing paradigm is a promising solution to high-resolution video analytics systems. 
The key lies in reducing redundant data and managing fluctuating inference workloads effectively. 
Previous work has focused on extracting regions of interest (RoIs) from videos and transmitting them to the cloud for processing. 
However, a naive Infrastructure as a Service (IaaS) resource configuration falls short in handling highly fluctuating workloads, leading to violations of Service Level Objectives (SLOs) and inefficient resource utilization.
Besides, these methods neglect the potential benefits of RoIs batching to leverage parallel processing.
In this work, we introduce Tangram, an efficient serverless cloud-edge video analytics system fully optimized for both communication and computation. 
Tangram adaptively aligns the RoIs into patches and transmits them to the scheduler in the cloud. 
The system employs a unique ``stitching'' method to batch the patches with various sizes from the edge cameras. 
Additionally, we develop an online SLO-aware batching algorithm that judiciously determines the optimal invoking time of the serverless function.
Experiments on our prototype reveal that Tangram can reduce bandwidth consumption and computation cost up to 74.30\% and 66.35\%, respectively, while maintaining SLO violations within 5\% and the accuracy loss negligible.
\end{abstract}


\section{Introduction}

High-resolution cameras are increasingly prevalent in various edge applications, e.g., surveillance~\cite{wang2020surveiledge}, traffic monitoring~\cite{li2023cross}, augmented reality~\cite{liu2019edge}, etc. 
High-resolution video analytics based on advanced computer vision models has become a vibrant research topic in recent years ~\cite{zhang2018awstream,wang2019bridging,vabus}. 

\begin{figure}[!t]
\begin{center}
\includegraphics[width=1\linewidth]{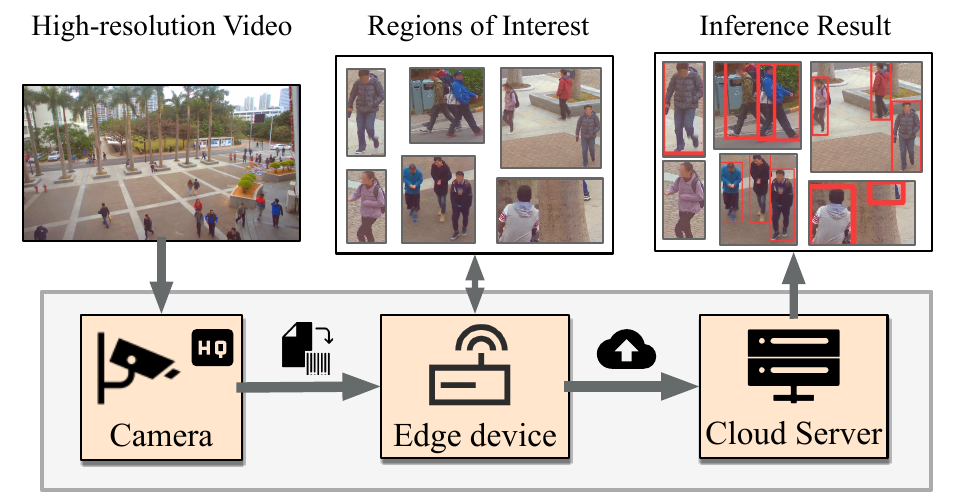}
\caption{A representative type of video analytics.}
\label{fig:VAP}
\vspace{-5mm}
\end{center}
\end{figure}

A straightforward way is to send videos to the cloud, which then executes deep neural network (DNN) model inference tasks and delivers useful visual feedback to users. 
In video analytics systems, Service Level Objectives (SLOs) refer to the total latency requirement from capturing the video to acquiring the model inference results, which is essential for real-time applications such as municipal surveillance and traffic management.
Unfortunately, transmitting high-resolution videos requires substantial network bandwidth resources. 
For example, transmitting a 4K video encoded in H.264 format at 30 frames per second typically requires a bandwidth of 13-34 Mbps~\cite{youtube}, which cannot be afforded by many edge devices. 
Subsequent research, as illustrated in Fig.~\ref{fig:VAP}, has identified considerable redundancy in high-resolution videos. 
As a solution, existing studies suggest transmitting only regions of interest (RoIs), thereby reducing bandwidth demands.
For example, server-driven approaches~\cite{zhang2022understanding,du2020server,9796853} allow edge devices to send low-quality videos to the cloud.
The cloud then identifies RoIs and provides feedback on their positions to edge devices. 
In the second transmission round, only these RoIs encoded in high quality are sent to the cloud.  
To avoid the two-round communication inherent in server-driven approaches, some content-aware work~\cite{vabus,xu2021approxnet,elf,chen2022context,eagleeye} have proposed to let edge devices identify RoIs independently.

\begin{figure}[!t]
    \centering
    \begin{subfigure}{0.36\textwidth}
        \centering
        \includegraphics[width=1\linewidth]{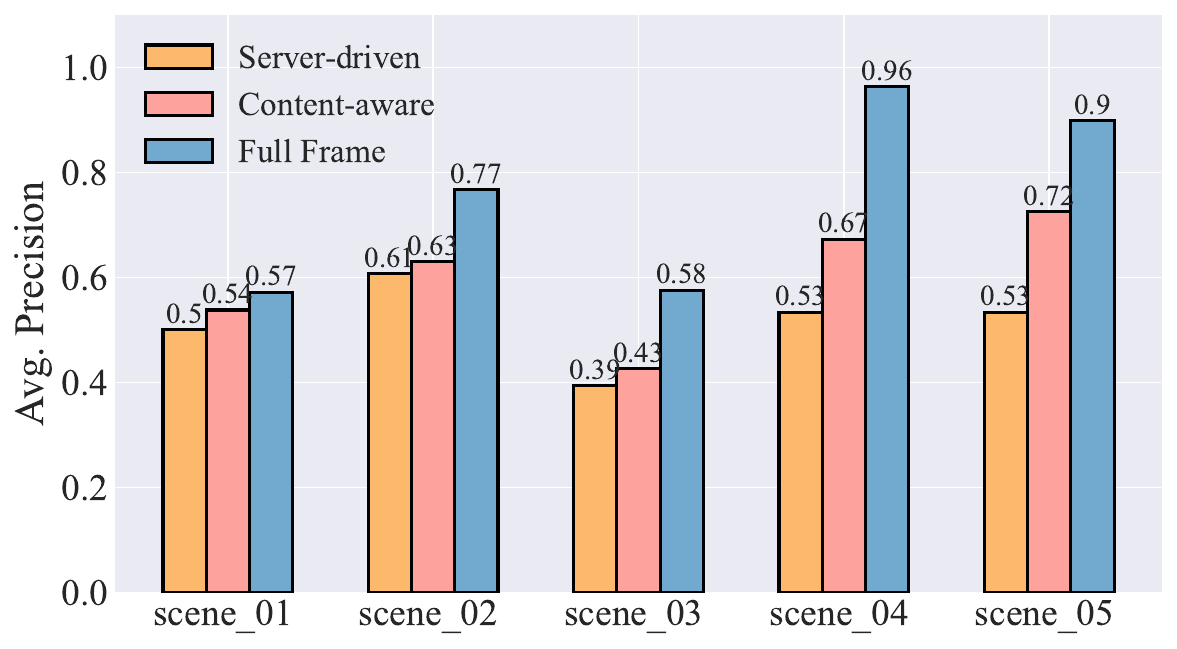}
        \caption{Loss of inference accuracy in high-resolution videos.}
        \label{fig1.5:sub1}
    \end{subfigure}
      \begin{subfigure}{0.12\textwidth}
        \centering
        \includegraphics[width=1\linewidth]{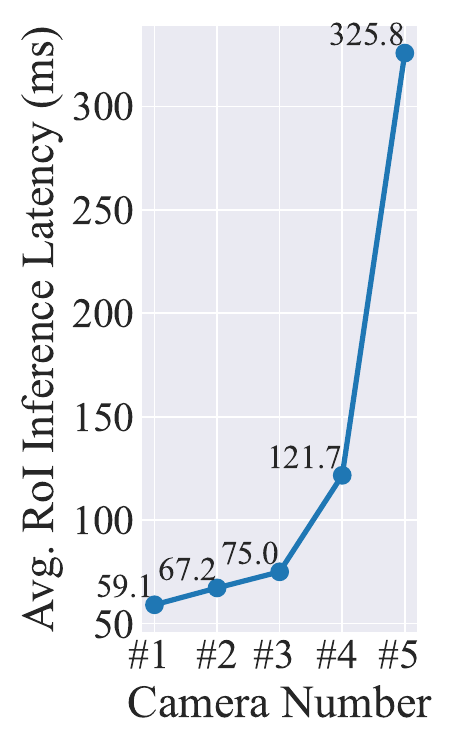}
        \caption{Latency v.s. \#Camera}
        \label{fig1.5:sub2}
    \end{subfigure}
    \caption{Previous methods are hard to adapt to high-resolution videos.}
    \label{fig1.5}
    \vspace{-2mm}
\end{figure}

We have conducted extensive experimental studies and found that most approaches cannot provide sufficient accuracy and throughput when handling high-resolution videos from many edge devices.
For instance, as shown in Fig~\ref{fig1.5}(a), we observed an average of 23.9\% and 14.1\% accuracy decline for server-driven and content-aware approaches in high-resolution object detection, respectively.
Besides, from Fig~\ref{fig1.5}(b), as the number of source cameras increases from 1 to 5, the average RoI inference time exponentially escalates from 59.07ms to 325.84ms with an NVIDIA GeForce RTX 4090 GPU.
%
%
As we study video analytics from the perspective of the end-to-end system, existing work has the following two weaknesses. 
First, although RoI-based methods can significantly improve communication efficiency by eliminating redundant contents, RoIs have different sizes, which complicates GPU inference that requires all inputs with the same size. 
A simple solution involving resizing or padding RoIs to unify their size so that they can be batched together, yet it reduces inference accuracy~\cite{ran2018deepdecision,dong2022collaborative,zhu2021dynamic} and adds additional computational burden~\cite{ali2022optimizing}. 
Second, the quantity and size of RoIs in video frames change dynamically, making the inference workload highly fluctuate~\cite{lu2022turbo}. 
If the provision of computing resources cannot keep pace with such dynamic workloads, it will result in severe response delay, potentially leading to breaches of SLOs.
Conversely, over-provisioned computing resources result in wastage, leading to substantial costs~\cite{elf,zhang2018awstream}. 

In this paper, we propose Tangram, an efficient cloud-edge video analytics system fully optimized for both communication and computation.
Tangram distinguishes itself from existing work through three innovative designs.
First, we design an adaptive frame partitioning algorithm to address the limitations of RoI extraction approaches in handling high-resolution videos. 
This lightweight filter can align the RoIs within the frame into \textit{patches} to mitigate the issue of object missing.
To facilitate efficient inference through batching, we propose to stitch several patches of different sizes to create a uniform \textit{canvas}. 
Different from resizing and padding, our method maintains inference accuracy and incurs minimal overhead. 
Second, we employ serverless functions to address fluctuating workloads. Unlike virtual machine instances that require a long time for launching and initialization, serverless functions can quickly scale up or down in tens of milliseconds~\cite{DBLP:conf/nsdi/FouladiWSBZBSPW17}. 
Moreover, users are only charged for their function execution time, typically measured in one-second units.
Such fine-grained auto-scaling ability and pricing strategy of serverless computing make it capable of tackling the fluctuating workloads in high-resolution video analytics. 
Although serverless functions have been studied in various applications, applying them for video analytics, particularly in combination with RoIs batching, remains an open challenge. 
Finally, we design a scheduler to decide how and when to feed the batches to the serverless functions to minimize the cost and SLO violation rate. 
%
%
We develop and deploy a prototype on a testbed running real video analytics workloads. 
Experimental results demonstrate that Tangram can reduce bandwidth consumption by up to 74.30\% and computation cost by up to 66.35\%, respectively, while maintaining SLO violations within 5\% and negligible accuracy loss.

The remainder of this paper is organized as follows. 
We present the motivation and challenge in Section~\ref{sec_II}. 
The design of Tangram is introduced in Section~\ref{sec_IV}, followed by the system implementation in Section~\ref{sec_V}. 
We conduct extensive experiments in Section~\ref{sec_VI}, and the related work is reviewed in Section~\ref{sec_VII}. 
Finally, we conclude this paper in Section~\ref{sec_VIII}.

\section{Motivation and Challenge}\label{sec_II}

In this section, we conduct experimental studies to investigate the issues present in real-world video analytics scenarios and discuss the motivation and challenge of this work.

\subsection{Redundancy in Video Inference Data}\label{sec_II_A}
High-resolution cameras capture videos consisting of a large number of objects (e.g., people and vehicles). 
Within each video frame, a small region containing objects is identified as an RoI. 
Conversely, the rest of the frame is dominated by the background (e.g., buildings and sky) and other irrelevant objects~\cite{remix}.
Table~\ref{table1} shows the redundancy of several real-world high-resolution videos. 
It is evident that RoIs constitute less than 10\% of most videos, and non-RoI computation overheads occupy up to 15.43\%. 
The primary reason is that high-resolution cameras usually have a larger field of view.
The redundancy in video analytics not only increases bandwidth consumption but also contributes to inefficiency in video inference. 
Therefore, extracting RoIs from videos and uploading them to the cloud for inference becomes a pivotal aspect of optimizing video analytics systems.

\begin{table}[!t]
\caption{Redundancy in video inference data on PANDA4K dataset\cite{wang2020panda}.}
\label{table1}
\begin{threeparttable}
\resizebox{0.48\textwidth}{!}{\begin{tabular}{@{}c|c|c|c|c@{}}
\toprule
\multicolumn{1}{l|}{\textbf{Index}} & \textbf{Scene Name (\# Frame)} & \textbf{\# Person} & \textbf{RoIs Prop\tnote{$\triangle$}.\quad (\%)} & \textbf{Redundancy\tnote{$\diamondsuit$} \ (\%)} \\ \midrule
1 & University Canteen (234) & 123 & 5.4510 & 12.39 \\
2 & OCT Habour (234) & 191 & 8.3141 & 11.28 \\
3 & Xili Crossroad (234) & 393 & 5.9132 & 9.24 \\
4 & Primary School (148) & 119 & 14.1561 & 15.43 \\
5 & Basketball Court (133) & 54 & 5.0354 & 15.43 \\
6 & Xinzhongguan (222) & 857 & 5.2316 & 10.93 \\
7 & University Campus (180) & 123 & 2.5860 & 10.31 \\
8 & Xili Street 1 (234) & 325 & 9.6297 & 10.65 \\
9 & Xili Street 2 (234) & 152 & 8.7498 & 9.25 \\
10 & Huaqiangbei (234) & 1730 & 9.6732 & 9.16 \\ \bottomrule
\end{tabular}}
 \begin{tablenotes}
        \footnotesize
        \item [\#] represents “ The number of ”;
        \item[$\triangle$] The ratio of the total area of RoIs to the whole frame;  
        \item[$\diamondsuit$] Non-RoIs inference time proportion. 
      \end{tablenotes}
    \end{threeparttable}
    \vspace{-6mm}
\end{table}

\subsection{Fluctuation of Inference Workloads}
\begin{figure}[!b]
    \centering
    \begin{subfigure}{0.48\textwidth}
        \centering
        \includegraphics[width=1\linewidth]{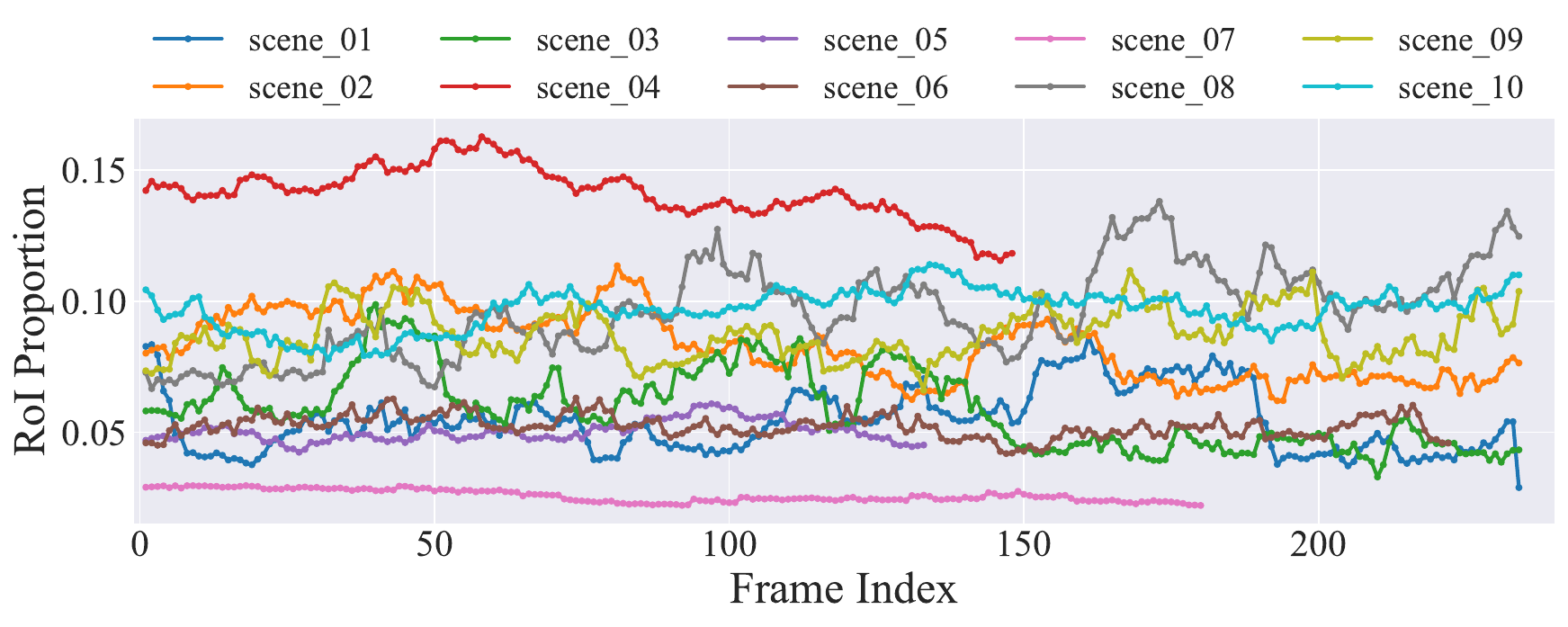}
        \caption{Temporal variation of the object area in ten scenes.}
        \label{fig1:sub1}
    \end{subfigure}
      \begin{subfigure}{0.48\textwidth}
        \centering
        \includegraphics[width=1\linewidth]{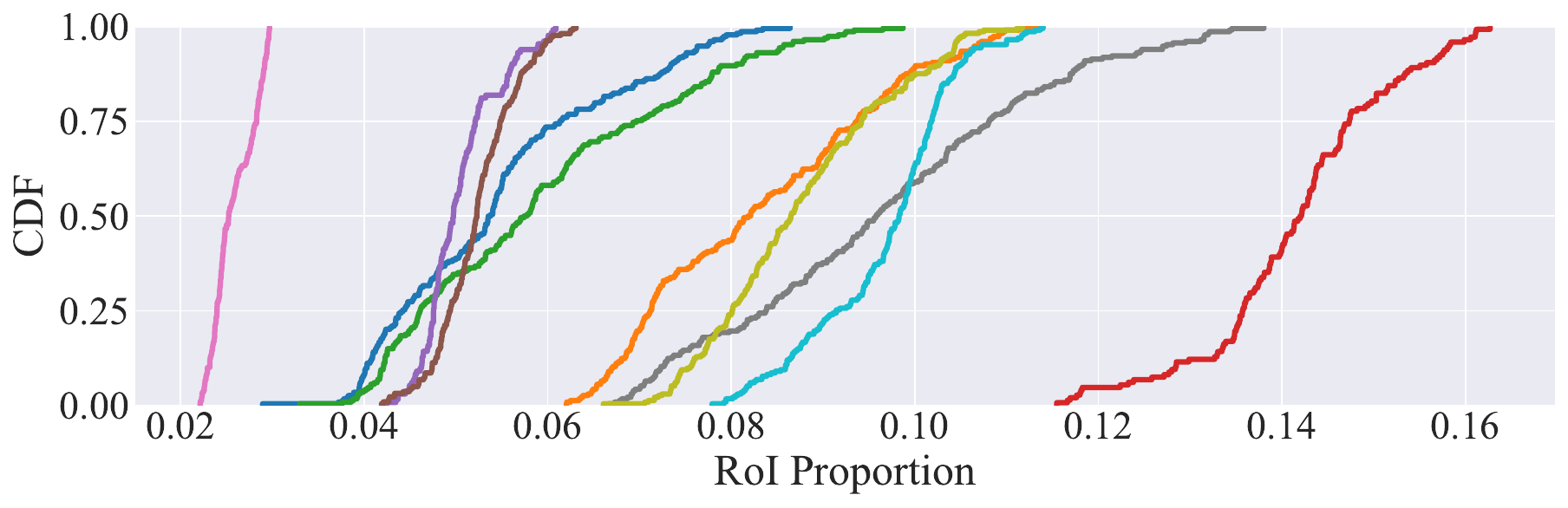}
        \caption{The cumulative distribution function~(CDF) of RoI proportion.}
        \label{fig1:sub2}
    \end{subfigure}
    \caption{The variation of video inference workloads in the ten real-world scenes.}
    \label{fig1}
\end{figure}
High-resolution cameras are commonly deployed in dynamic scenes such as traffic intersections, building entrances, and pedestrian streets, where the quantity and size of RoIs change frequently. 
We further conduct a deeper investigation of the variation of video inference workloads. 
Fig.~\ref{fig1}(a) illustrates the proportion of the RoIs in each frame varying over time, and Fig.~\ref{fig1}(b) depicts the distribution of RoI areas within each video. 
Typically, in most scenes, the RoIs fluctuate within a range of 5\% to 15\%. 
Peaks usually appear irregularly in these videos.
It can be observed that these fluctuations do not follow any predictable patterns or rules. 
Traditional virtual machines are less optimal for such dynamic scenarios due to their slower startup time.
Moreover, maintaining over-provisioned resources would result in unnecessary wastage. 

\subsection{Challenges of RoIs Batching}\label{motivation3}

\begin{figure}[!t]
    \centering
    \begin{subfigure}{0.24\textwidth}
        \centering
        \includegraphics[width=1\linewidth]{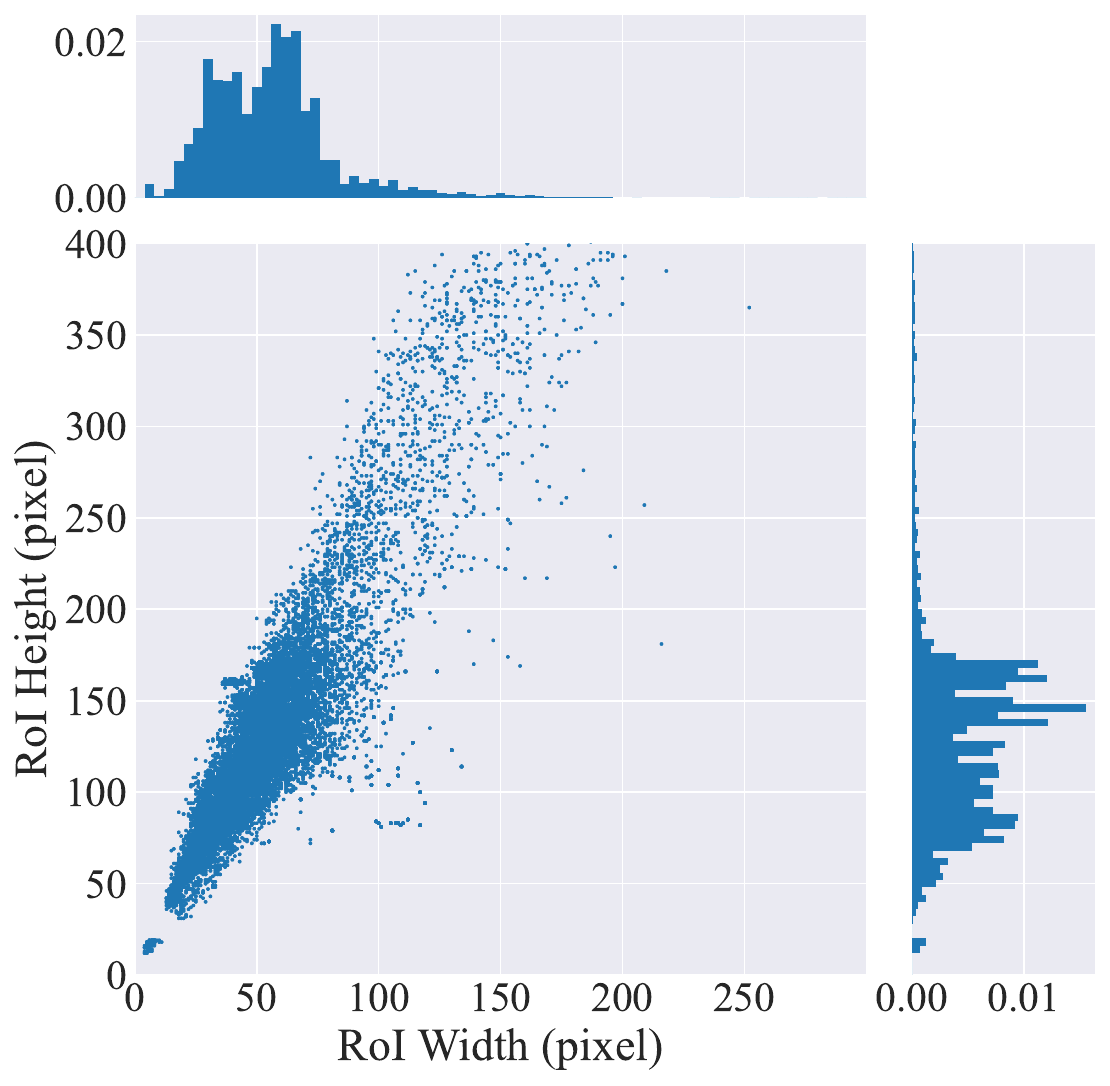}
        \caption{Sizes of RoIs in scene\_01.}
        \label{fig2:sub1}
    \end{subfigure}
      \begin{subfigure}{0.24\textwidth}
        \centering
        \includegraphics[width=1\linewidth]{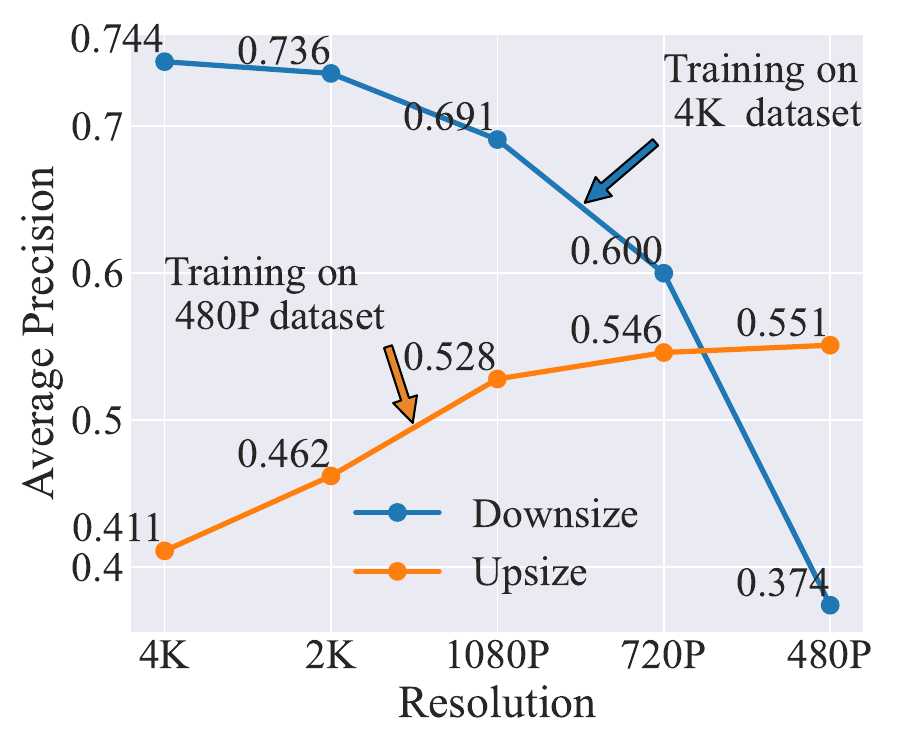}
        \caption{The inference accuracy of the PANDA dataset at different resolutions on Yolov8x.}
        \label{fig2:sub2}
    \end{subfigure}
    \caption{Challenges of RoIs batching.}
    \label{fig2}
    \vspace{-2mm}
\end{figure}

Batching is a recognized and effective technique to enhance inference efficiency in video analytics \cite{crankshaw2017clipper, Zhang2020EnablingCS}.
In order to batch requests and feed them to the model service, the input size needs to be the same.
However, as shown in Fig.~\ref{fig2}(a), the size of RoIs in high-resolution videos varies greatly, which makes it difficult to batch them together.   
A common approach is to batch such RoIs by resizing or padding.
To evaluate their efficiency, we trained two Yolov8x models, one adapted for 4K and the other for 480P resolution.
As shown in Fig.~\ref{fig2}(b), 480p and 4K models are fed upsized RoIs (orange) and downsized RoIs (blue), respectively. We observe a noticeable drop in accuracy when the input size does not match the model.
Besides, adopting a padding approach would inevitably induce extra computational resources.
Therefore, finding a method to batch RoIs of various sizes efficiently without compromising the accuracy of video inference is very challenging.

\section{Tangram Design}\label{sec_IV}
Tangram is a cloud-edge video analytics system that leverages serverless computing for high-resolution video analytics. 
It can not only reduce the bandwidth consumption but also minimize the cost of function invocations while satisfying the SLO.

\begin{figure*}[htbp]
\begin{center}
\includegraphics[width=1\linewidth]{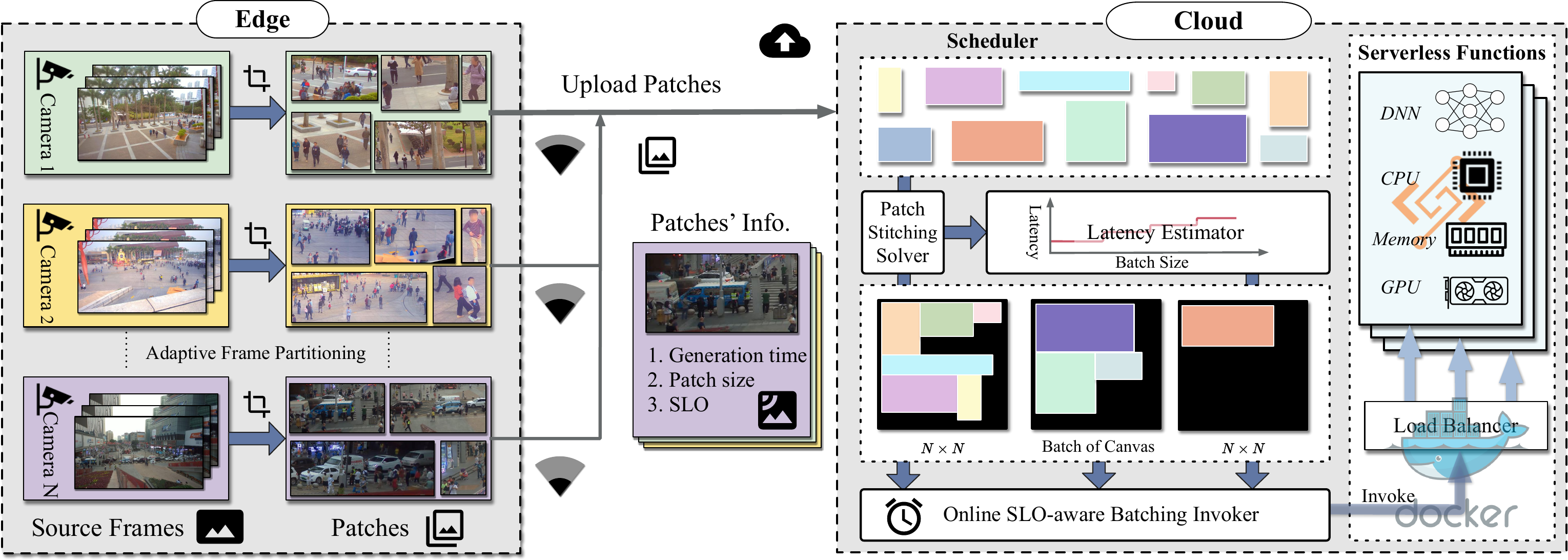}
\caption{Overview of Tangram}
\label{fig:Overview}
\end{center}
\vspace{-4mm}
\end{figure*}

As shown in Fig.~\ref{fig:Overview}, Tangram consists of two primary components: the edge and the cloud server. The cameras capture video at the edge and run the adaptive frame partitioning algorithm in real time.
Based on the dynamic characteristics of the objects, the RoIs within the collected video frames are aligned into patches of various sizes.
The edge then uploads all the patches and their additional information to the cloud, including the generation time, the patch's size, and SLO. 
Subsequently, in the cloud \textit{Scheduler}, the \textit{Patch-stitching Solver} stitches all the patches together to form a batch of uniform-size canvases. 
%
Meanwhile, the \textit{Latency Estimator} is responsible for estimating the inference time of a batch of canvases and alerting \textit{Online SLO-aware Batching Invoker} when to trigger the inference, i.e., dispatching the batch of canvases for processing by serverless function. 

\subsection{Adaptive Frame Partitioning}\label{Adaptive}
High-resolution cameras are usually deployed with fixed positions and viewing angles.
The background modeling (e.g., Gaussian mixture model \cite{stauffer1999adaptive}) can segment foreground objects and exclude static background, which is well-suited for RoI extraction.
We also compared other models in Section \ref{accuracy}.
%
However, due to the tiny area (about $50\times50$ pixels) of some distant objects in the high-resolution video, many small objects failed to be detected by traditional background modeling algorithms. 
To improve the recall of objects, we propose an adaptive frame partitioning approach to reserve all the small foreground objects as much as possible. 
The insight is that more objects could be found near or between regions with a high occurrence of foreground objects~\cite{remix}.
The pseudo-code is shown in Algorithm \ref{alg_apfa}, which contains the following main steps.

\begin{figure}[htbp]
\begin{center}
\includegraphics[width=1\linewidth]{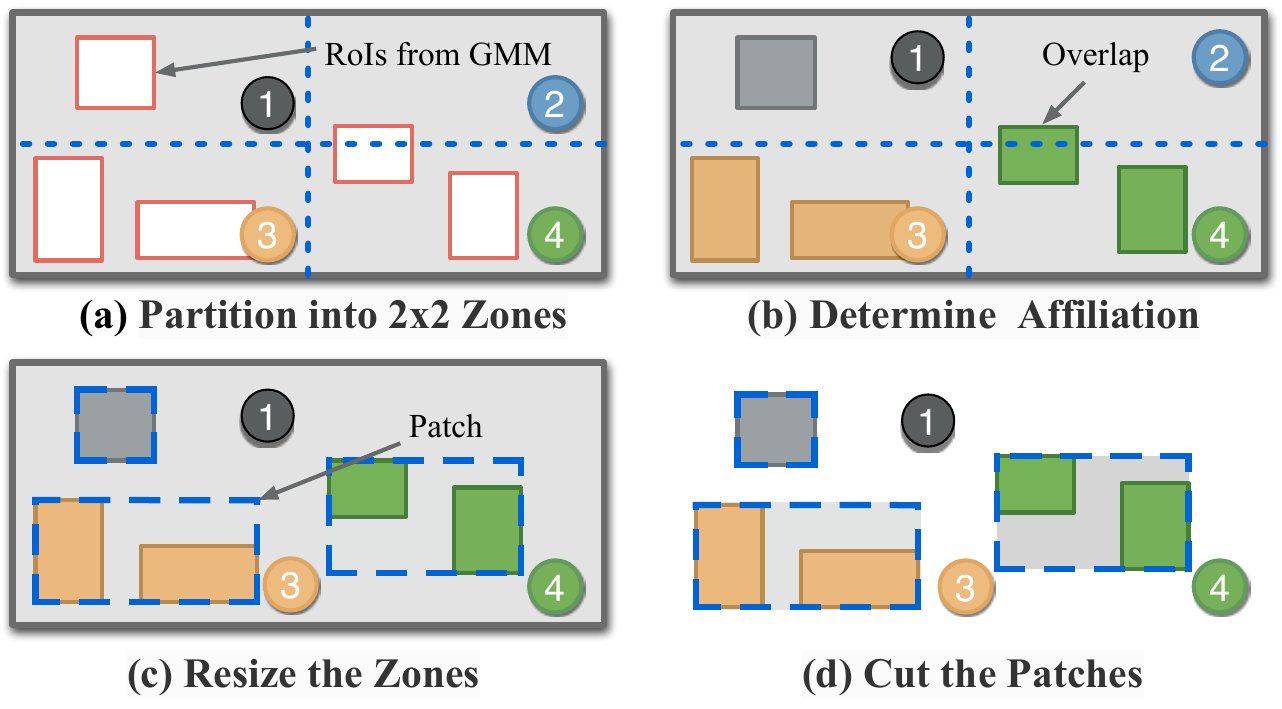}
\caption{The process of adaptive frame partitioning algorithm.}
\label{fig:Adaptive}
\end{center}
\vspace{-3mm}
\end{figure}

\begin{algorithm}[!t] 
	\caption{\label{alg_apfa}Adaptive Frame Partitioning Algorithm}
	\LinesNumbered 
	\KwIn{Source frame's resolution: $W\times H$; Zone shape: $X\times Y$;  RoIs $\mathbb{B}=\{1,2,\ldots,B\}$ from GMM.} 
	\KwOut{Patches $ \mathbb{I}$.}
	Divide the frame into $X\times Y$ zones $\mathbb{R
}=\{1,2,\ldots,{X\times Y}\}$, each zone has the same size of $\frac{W}{X}\times\frac{H}{Y}$\label{line1}\;
	Set $\mathbb{L}_{r}=\{\emptyset\}$ for every zone $r\in\mathbb{R}$\;
	\For{$b \in \mathbb{B}$}{
		\For {$r \in \mathbb{R}$}{
		$S_{b,r}\leftarrow \text{Overlap\_area}(b,r)$\;
		}
		$r^{*} \leftarrow \arg \max_{r\in R}\{S_{b,r}\}$\;
		$\mathbb L_{r^{*}}.$append$(b)$\;
	}
	\For{$r \in \mathbb{R}$}{
	\If {$\mathbb{L}_{r}\neq\{\emptyset\}$}{
		Resize each zone to the minimum enclosing rectangle that covers all the RoIs in $\mathbb L_{r}$\;
		Cut the zone as the patch and append it to  $\mathbb{I}$\;
		}
	}
\end{algorithm}

\begin{enumerate}
\item \textbf{Generate RoIs:} Each video frame is evenly divided into $X\times Y$ zones. Fig.~\ref{fig:Adaptive} shows an example when $X=Y=2$. We then use the Gaussian mixture model (GMM) \cite{stauffer1999adaptive} to obtain the RoIs.
\item \textbf{Determine affiliation:} Each RoI is associated with a specific zone (Fig.~\ref{fig:Adaptive}(b)). For every RoI $b$, we calculate the overlap area $S_{b,r}$ with each zone $r$. The RoI $b$ is assigned to the zone $r^{*}$ with the maximum overlap area, and it is added to the corresponding zone's list $\mathbb{L}_{r^{*}}$ (Lines 3-9).
\item \textbf{{Resize the zones}:} 
We resize each zone to the minimum enclosing rectangle that covers all RoIs associated with it (Fig.~\ref{fig:Adaptive}(c), Lines 10-12).
\item \textbf{Cut the patches}: Finally, each zone is cut out to form a patch (Line 13). It's worth noting that all the patches belonging to the same frame have the same SLO.
\end{enumerate}

\subsection{Batching Problem Description}
Since the patches are of different sizes, it is challenging to batch them together.
To tackle this issue, we employ a fixed-size rectangle canvas to hold the patches. 
When the canvas is full and cannot accommodate more patches, a new canvas will be opened to accommodate the patches. 
At an appropriate time, multiple canvases can be batched together for serverless function execution. 
Our goal is to minimize the computing cost of video inference while meeting the SLO on a serverless platform.
%

In this paper, we use Alibaba Cloud Function Compute~\cite{ali_fc}, a serverless computing platform with GPU instance support, as the cost model of serverless function. 
An invocation of serverless function is charged based on the execution time and the allocated resource as~\cite{ali} 
\begin{equation}\label{eq1}
C_{Ali} = T_{f}\cdot \left(n_{C}\cdot P_{C}+m_{M}\cdot P_{M}+m_{G}\cdot P_{G} \right)+ P_{req},
\end{equation}
where $T_{f}$ is the function execution time, $n_{C}$, $m_{M}$, and $m_{G}$ are the vCPU, GB of memory, and GB of GPU memory used by the function instance, respectively. 
The $P_{C}$ (i.e., $2.138\times10^{-5}\$/vCPU\cdot s$ ), $P_{M}$ (i.e., $2.138\times10^{-5}\$/GB\cdot s$), and $P_{G}$ (i.e., $1.05\times10^{-4}\$/GB\cdot s$) are the unit price of vCPU, memory, and GPU memory, respectively, $P_{req}$ (i.e., $2\times10^{-7}$\$) is the basic cost of each invocation. 

Let $\mathbb I=\{1,\ldots,I\}$ denote the set of patches, $\mathbb J=\{1,\ldots,J\}$ denote the set of canvases, and ${\mathbb K}=\{1,\ldots,K\}$ denote the set of batches.
We define a binary variable~$x^{j}_{i}$, where~$x^{j}_{i} = 1$ if patch $i$ is in canvas $j$, otherwise~$x^{j}_{i} = 0$. 
The~$y^{k}_{j}=1$ indicates that canvas $j$ is placed in batch $k$, and is $0$ otherwise. 
And~$z^{k}_{i} =1$ denotes that patch $i$ is in batch $k$, else it is $0$. 
Our objective is to minimize the total computation cost of patch inference, which is
\begin{align}
& {\min}
& & \sum^{K}_{k=1}T^{k}_{f} \left(n_{C}\cdot P_{C}+m_{M}\cdot P_{M}+m_{G}\cdot P_{G} \right)+ P_{req} \label{nphard}\\
& \text{s.t.}
& & \sum^{J}_{j=1}  x^{j}_{i} =1, \sum^{K}_{k=1}  z^{k}_{i} =1, \forall i \in\mathbb I,  \label{st1}\\
& & & \sum^{I}_{i=1} s_i  x^{j}_{i} \leq S ,  \forall j\in\mathbb J,\label{st2} \\
 & & & w\sum^{J}_{j=1}y^{k}_{j}+ \tau \leq m_{G}, \forall k\in\mathbb K, \label{st2.5}\\
& & & T_{i,wait}+T^{k}_{f}\leq SLO_i, i \in \{i| z_i^k=1,\forall i\in\mathbb I\},\label{st3}\\
& & & T^{k}_{f} =  f (\sum^{J}_{j=1}y^{k}_{j},n^{k}_{C},m^{k}_{m},m^{k}_{G}), \forall k\in\mathbb K, \label{st6}
\end{align}
where $\tau$ is the model size, $w$ represents the GPU memory occupied by a single canvas, $s_{i}$ is the size of patch $i$, and $S$ is the canvas size. 

Constraint~(\ref{st1}) states that each patch can only be placed on a particular canvas in a specific batch. 
Constraint~(\ref{st2}) implies that the total area of all patches in a canvas should not exceed the canvas' area. 
Constraint~(\ref{st2.5}) specifies that the GPU memory usage of each batch should not exceed the resource allocated to the function. 
Constraint~(\ref{st3}) asserts that each patch should not violate the SLO, where $T_{i,wait}$ and $SLO_{i}$ are the waiting time and SLO of patch $i$. 
Constraint (\ref{st6}) is the inference time of batch $k$, which is related to the size of the batch and the function configuration.

\subsection{Algorithm Design}
To address the challenges of configuring online batch processing for DNN inference mentioned in Section~\ref{motivation3}, we design a novel SLO-aware batching algorithm. 
This algorithm eliminates intricate batching parameter design and automatically invokes the serverless function according to the SLO.
It stitches the patches onto a sequence of fixed-size canvases (e.g., $1024\times 1024$) as many as possible. 
The advantages of this approach are twofold: 1) it fully utilizes the benefits of batch processing; 2) our method does not require patch resizing, thus avoiding information loss.

In fact, over a continuous period, the scheduler receives patches one after another, and we only need to determine when to stop waiting and invoke the function. 
That is the core idea of our scheduler, which comprises the following three modules. 

\textbf{Online SLO-aware Batching Invoker.} SLO-aware batching invoker continuously monitors the current canvases and calculates the \textit{remaining time} $t_{remain}$. 
Once current time aligns $t_{remain}$, it immediately batches all current canvases and triggers the function execution.
The details of the SLO-aware batching algorithm are described in Algorithm \ref{SLO-aware}. 

The edge sends the patch $i$ and its information $\mathbb P_{i}$, including the width $w_{i}$, height $h_{i}$, and deadline $t_{ddl_{i}}$ (i.e., the generation time plus the SLO). 
The scheduler initializes a set of blank canvases $\mathbb C$ with the size of $M\times N$.
Once the cloud receives a patch, it performs the following operations.
\begin{enumerate}
\item Push the patch into the queue $\mathbb Q$ and adopt the earliest deadline among all patches in $\mathbb Q$ as the deadline $t_{DDL}$. 
Save the old canvas set $\mathbb C_{old}$ (Lines 4-7).

\item According to the current queue $\mathbb Q$ and canvas size $M\times N$, the Patch-stitching Solver stitches all the patches to the canvases (Line~8). 
After that, the Latency Estimator gives the conservative inference time (i.e., $T_{slack}$) of the canvases $\mathbb C$ (Line 9). 
Then the $t_{remain}$ is calculated by
\begin{equation}\label{remain}
t_{remain} = T_{DDL} - T_{slack}.
\end{equation}

\item Once current time aligns $t_{remain}$, all current canvases $\mathbb C$ should be invoked for function execution immediately (Lines 19-22).

\item If the estimated $t_{remain}$ has already exceeded the current time, it means that adding this patch to the queue $\mathbb Q$ would violate the SLO. 
Besides, when the memory occupied by the number of canvases exceeds the GPU memory of the function instance, the patch should form a new queue. 
Meanwhile, the old canvas set $\mathbb C_{old}$ should be executed immediately (Lines 11-17) in both situations.
\end{enumerate}

\textbf{Latency Estimator.} It is necessary to approximate the inference time required for different batch sizes.
In this module, we try to get a relatively conservative time $T_{slack}$, which can minimize the violation rate of the SLO. 
Specifically, canvases of size $M\times N$ featuring diverse patch compositions are grouped into different batch sizes. 
Each group undergoes 1000 inference iterations, with their corresponding average time $\mu_{M\times N}$ and standard deviation $\sigma_{M\times N}$ being recorded.  
The objective is to harness the Law of Large Numbers to attain relatively precise and feasible estimations~\cite{yu2022orloj,park20183sigma}.
Therefore, we set the \textit{slack time} $T_{slack}$ as the mean value plus three times the standard deviation, which is 
\begin{equation}
T_{slack} = \mu_{M\times N} + 3\cdot\sigma_{M\times N}.
\end{equation}
This conservative estimation allows the function to have sufficient time for inference without violating the SLO. 
Notably, the Latency Estimator is profiled in the offline stage, so its cost and latency can be ignored.

\textbf{Patch-stitching Solver.} This module is tasked with stitching the existing patches onto the canvas together.
%
In our case, the patch cannot be overlapped, rotated, resized, or padded. 
The pseudo-code for the Patch-stitching Solver is delineated in Algorithm \ref{SLO-aware} (Lines 24-39).
Specifically, Patch-stitching solver selects a rectangular space $c$ that can contain the patch $i$ (i.e, $w_c \geq w_i$ and $h_c \geq h_i$) and has the smallest $\min \left(w_c-w_i, h_c-h_i\right)$. 
Then, it places the patch $i$ on the bottom-left corner of rectangle $c$ (Line 31).
Next, the residual space is divided into two non-overlapping rectangles, $c\prime$ and $c\prime\prime$, with the division based on the shorter side (Lines 32-33). 
This process continues until no free space can accommodate the next patch. Otherwise, the solver restarts with a new blank canvas (Line 36).
%
%

\begin{algorithm}[!t]  
	\caption{SLO-aware Batching Algorithm}\label{SLO-aware}
	\LinesNumbered 
	\KwIn{The information $\mathbb{P}_{i}=\{w_{i},h_{i},t_{ddl_{i}}\}$ of patch $i$, Canvas size $M\times N$}
	Initialize a queue $\mathbb{Q}=\{\emptyset\}$ to save the patches' info\;
	$\mathbb C\leftarrow \{\emptyset\}$, $\mathbb{C}_{old}\leftarrow \{\emptyset\}$\;
        \SetKwProg{Fn}{Function}{:}{end}	
 \While{True}{
		\If{\text{received patch} $i$ \text{with} $\mathbb{P}_{i}$}{
			$\mathbb Q.$append$(\mathbb P_{i})$\;
			$t_{DDL} \leftarrow \min\{t_{ddl_i}\}_{{\mathbb P}_i\in\mathbb Q}$\;
			$\mathbb{C}_{old} \leftarrow \mathbb{C}$\;
			$\mathbb C \leftarrow \textit{Patch\_stitching\_solver}(\mathbb Q,M,N)$\;
			$T_{slack}\leftarrow\textit{Latency\_estimator}(\mathbb C)$\;
			$t_{remain}\leftarrow t_{DDL} - T_{slack}$\;
			\If{$t_{remain}> t$ or memory$(\mathbb{C})>m_{G}-\tau$ }{
			Invoke$(\mathbb{C}_{old})$\; 
			$\mathbb{Q}\leftarrow \{\mathbb{P}_{i}\}$, $\mathbb{C}_{old}\leftarrow \{\emptyset\}$\;
			$\mathbb C \leftarrow \textit{Patch\_stitching\_solver}(\mathbb Q,M,N)$\;
			$T_{slack}\leftarrow\textit{Latency\_estimator}(\mathbb C)$\;
			$t_{remain}\leftarrow t_{DDL} - T_{slack}$\;
			}
		}
		\If{$t=T_{remain}$}{
			Invoke$(\mathbb C)$\;
			$\mathbb{Q}\leftarrow\{\emptyset\}$, $\mathbb C\leftarrow \{\emptyset\}$, $\mathbb{C}_{old}\leftarrow \{\emptyset\}$\;
		}
	}
\Fn{Patch\_stitching\_solver($\mathbb{Q},M,N$)}{
$C =$ \{$(M,N)$\}\;
\For{patch $i \in \mathbb{Q}$}{
$C_p=\left\{c \in C \mid\left(w_c \geq w_i\right) \cap\left(h_c \geq h_i\right)\right\}$\;
\If{$C_p\neq\emptyset$}{
Decide the $c\in C_p$ to stitch the patch onto\;
$c\leftarrow \arg\min_c\left(\min(w_c-w_i, h_c-h_i)\right)$\;
Place the patch $i$ at the bottom-left of $c$\;
Split $c$ into $c^{\prime}$ and $c^{\prime\prime}$ on a shorter axis\;
Set $C=C \cup\left\{c^{\prime}, c^{\prime \prime}\right\} \backslash c$\;
}
\Else{
Re-initialize a new canvas $C$\;
}
}
}
\end{algorithm}

\begin{figure}[!t]
\begin{center}
\includegraphics[width=1\linewidth]{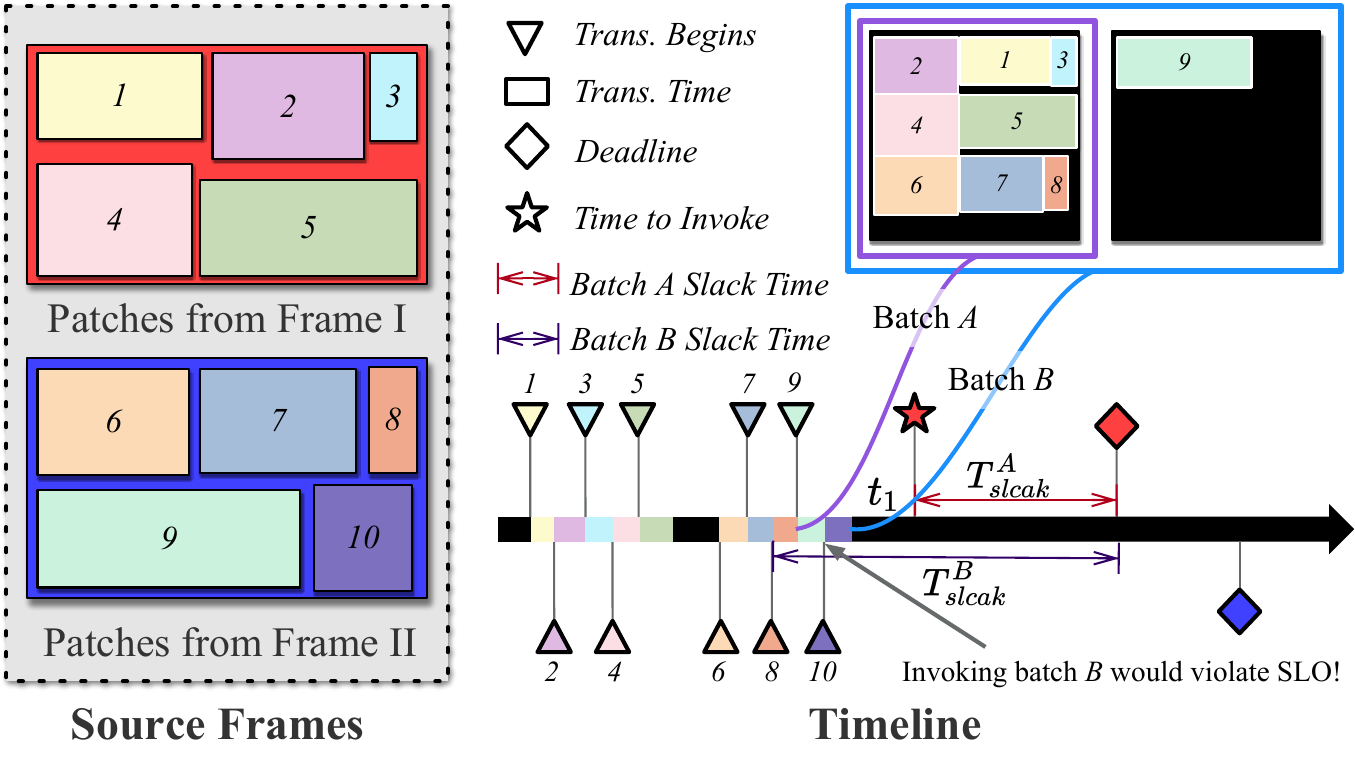}
\caption{A example of SLO-aware batching algorithm.}
\label{fig:SLO-aware}
\end{center}
\end{figure}

Fig.~\ref{fig:SLO-aware} shows a representative example of our pipelines. Suppose there are two source frames from the cameras and patches \textit{1} to \textit{5} and \textit{6} to \textit{10} are the outcomes from the adaptive frame partitioning algorithm (i.e., Algorithm~\ref{alg_apfa}) of the frame I and frame II, respectively.

The timeline on the right side of the figure illustrates the algorithm's progression throughout its execution. Triangles mark the initiation of patch transmission, and colored blocks denote the duration of the transmission. The red and blue rhombus indicate the deadline for batch \textit{A} and \textit{B}, respectively. Furthermore, the spans highlighted by the red and purple double arrows represent the estimated slack time for batch \textit{A} and \textit{B}, respectively.

Initially, patches \textit{1} to \textit{7} are transmitted in sequence. 
With each patch's arrival, the scheduler activates the patch stitching solver to get the current canvas $\mathbb C$ (Line 8, as shown in the upper right corner of Fig.~\ref{fig:SLO-aware}). 
Upon the arrival of patch \textit{8}, all existing patches can be accommodated on a single canvas. 
Meanwhile, the latency estimator calculates the slack time $T^{A}_{slack}$ of batch \textit{A} (Line 9) and determines the remaining time $t_{remain}$ to the deadline (line 10). 
Consequently, this canvas must be invoked before $t_{1}$ (marked by the red star) to ensure adherence to the SLO. 
Next, as patch \textit{9} arrives, the patch stitching solver cannot stitch it on the existing canvas. 
This change causes the slack time estimated by the latency estimator to shift from $T_{slack}^{A}$ to $T_{slack}^{B}$, surpassing the current time. 
As a result, the invoker immediately dispatches the first canvas (i.e., Batch \textit{A}) for function execution, leaving patch \textit{9} to form part of the next canvas.

\section{Implementation}\label{sec_V}
We develop a prototype of Tangram in Python and C++. 
We conduct our experiments on our cloud server with an Intel(R) Xeon(R) Gold 6326 CPU, 128 GB of RAM, and 2 NVIDIA GeForce RTX 4090 GPUs with 24 GB of VRAM and use NVIDIA Jetson Nano  4GB as the edge device. 
The cloud server and edge device are connected with the TP-LINK TL-WDR5620. 
The operating system of the server and edge device are both Ubuntu 20.04.6 LTS. 

In the software setup, we build our patch extraction algorithm on top of the cuda::BackgroundSubtractorMOG2 implemented by OpenCV~\cite{opencv} on Jetson. 
For serverless function, we use NVIDIA docker~\cite{nvidia} to run the DNN model on our cloud server and utilize FastAPI~\cite{fastapi} as the web framework and NGINX~\cite{nginx} as the load balancer. 
Yolov8x model serving based on pytorch is modified from official implementation~\cite{Jocher_YOLO_by_Ultralytics_2023}. 
The edge device and cloud server are connected through the HTTP protocol.

Our Tangram system operates orthogonally to the DNN model and RoI extraction algorithms, making it flexible to the downstream tasks of video analytics (e.g., keypoint detection or segmentation). 
Specifically, the lightweight adaptive frame partitioning algorithm is implemented by API on the edge device:

{\ttfamily def partition(Frame,X,Y,M,N)->List[Patch]},
which divides the {\ttfamily Frame}, sized {\ttfamily M$\times$N}, into {\ttfamily X$\times$Y} zones and obtains a list of patches and their generation time, sizes, and SLO. 
In addition, the {\ttfamily X} and {\ttfamily Y} are utilized to control the granularity of the partitioning. 
Tangram can be initialized from the instance in the cloud: {\ttfamily class Tangram(canvas\_size: List)}, where the {\ttfamily canvas\_size} can be experientially determined based on the camera's resolution. 
Next, we need to implement the following two APIs:  

{\ttfamily 1. def receive\_patch(patch: numpy.array)}

{\ttfamily 2. def invoke(canvases: numpy.array)}

Tangram employs the first API to receive the patch and its information and the second API to invoke one inference to the serverless function for a batch of canvases. 
The plug-and-play design of Tangram obviates the need for modifications to the original cloud-edge system, and replacing the components can be adapted to other scenarios. 
For instance, if we expect an analysis of pedestrian action, we only need to replace the serverless function with a pose estimation model.

\section{Evaluation}\label{sec_VI}
In this section, we first describe the experimental setting and then validate the effectiveness of the adaptive frame partitioning algorithm using Alibaba Cloud Function Compute~\cite{ali_fc}, a public serverless platform. 
Finally, we evaluate the Tangram in an end-to-end video analytics scenario with SLO restriction on our testbed and report its performance.

\subsection{Experimental Setting}
\begin{figure*}[htbp]
\begin{center}
\includegraphics[width=1\linewidth]{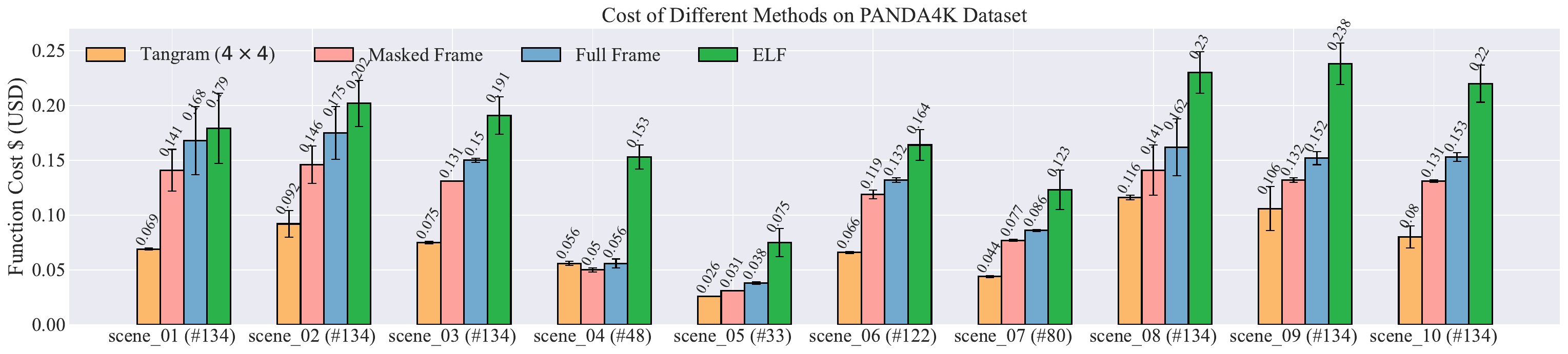}
\caption{Cost of Tangram, ELF, Masked Frame, and Full Frame on ten scenes of PANDA4K (\# the number of evaluation frames) on Alibaba Cloud Function Compute.}
\label{fig:experiment1}
\end{center}
\end{figure*}
\begin{figure*}[htbp]
\begin{center}
\includegraphics[width=1\linewidth]{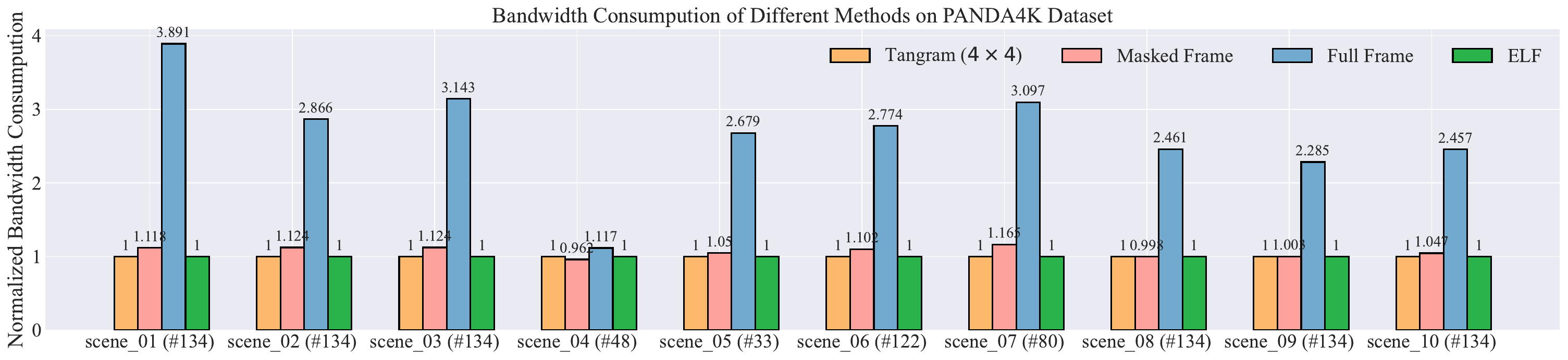}
\caption{Bandwidth Consumption of Tangram, ELF, Masked Frame, and Full Frame on ten scenes of PANDA4K (\# the number of evaluation frames) on Alibaba Cloud Function Compute.}
\label{fig:experiment2}
\end{center}
\end{figure*}


We consider the object detection DNN model Yolov8x\cite{Jocher_YOLO_by_Ultralytics_2023} with 68.2M parameters. 
We use the PANDA\cite{wang2020panda} video sequences, a high-resolution human-centric video dataset for pedestrains detection captured by a stationary gigapixel camera. 
The original training dataset has ten scenes, including 2087 frames of $26753\times15052$ resolution. 
We resize the frames to $3840\times2160$ (4K) as the PANDA4K dataset. 
Specifically, we combine the first 100 frames from each scene to form a training set of 1000 samples. 
The remaining frames are used for evaluation.

In our experiment, the specifications of the serverless function are two cores vCPU, 4GB memory, and 6GB GPU memory. 
Furthermore, the concurrency of each function is set to 1. NGINX employs a default load-balancing method. 
The cost of the function invocation is calculated by Eqn.~(\ref{eq1}). 
Unless otherwise specified, the size of the canvases in this paper is set to $M=N=1024$. 
We compare Tangram with the other state-of-the-arts.
\begin{itemize}
\item \textit{Full Frame}: It directly transmits the original frames at 4K resolution to the scheduler and triggers the function in sequence (each frame as a single request).
\item \textit{Masked Frame}~\cite{liu2022adamask}: The non-RoIs in the original frame are masked. It only transmits the masked frame at a 4K resolution and triggers in sequence (each frame as a single request).
\item \textit{ELF}\cite{elf}: All patches are cut out, transmitted to the cloud, and triggered in sequence.
\item \textit{Clipper}\cite{crankshaw2017clipper}: We implement the dynamic batch size strategy in \cite{crankshaw2017clipper}, a variant of Additive-Increase, Multiplicative-Decrease schemes.
\item \textit{MArk}\cite{Zhang2020EnablingCS}: A strategy that jointly takes into account batch size and timeout. We set an appropriate timeout for each bandwidth setting. 
\end{itemize}

\subsection{Performance of Tangram}\label{experiment1}

We first validate our approach by employing the adaptive frame partitioning algorithm (with $4\times4$ zones) to every frame and stitching those patches onto the canvases as a single request, denoted as Tangram $4\times4$. 
Fig.~\ref{fig:experiment1} shows the cost of serverless function execution of different methods on ten scenes of the PANDA4K dataset. 
The Tangram performs best in almost all scenarios by reducing the cost to 66.42\%, 57.39\%, and 41.13\% compared with Masked Frame, Full Frame, and ELF on average. 
Fig.~\ref{fig:experiment2} shows the normalized bandwidth consumption of different approaches. 
As we can see, by employing the adaptive frame partitioning algorithm, we remove the non-RoIs from the original video frames, reducing the bandwidth consumption compared to the Full Frame approach. 
Specifically, the reduction varies between 10.47\% to 74.30\% in ten scenes. 
The impact of different partition parameters on bandwidth is demonstrated in Table~\ref{table2}, we find that more fine-grained zone divisions can save more bandwidth.
%


\begin{table}[!t]
\caption{Bandwidth Consumption Normalized to the Full Frame Approach on PANDA4K dataset.}
\centering
\label{table2}
\begin{tabular}{@{}c|c|c|c@{}}
\toprule
\multicolumn{1}{c|}{\multirow{2}{*}{\textbf{Scene Index}}} &
  \multicolumn{3}{c}{\textbf{Configuration}} \\ \cmidrule(l){2-4} 
\multicolumn{1}{c|}{} &
  \multicolumn{1}{l|}{\textbf{2x2 (\%)}} &
  \multicolumn{1}{l|}{\textbf{4x4 (\%)}} &
  \multicolumn{1}{l}{\textbf{6x6 (\%)}} \\ \midrule
\textbf{scene\_01} & 44.2 & 25.7 & 19.3 \\
\textbf{scene\_02} & 45.6 & 34.9 & 29.2 \\
\textbf{scene\_03} & 56.2 & 31.8 & 25.6 \\
\textbf{scene\_04} & 89.7 & 89.5 & 50.3 \\
\textbf{scene\_05} & 95.4 & 37.3 & 25.7 \\
\textbf{scene\_06} & 49.8 & 36.1 & 30.1 \\
\textbf{scene\_07} & 52.3 & 32.3 & 32.3 \\
\textbf{scene\_08} & 58.3 & 40.6 & 30.7 \\
\textbf{scene\_09} & 58.9 & 43.8 & 35.9 \\
\textbf{scene\_10} & 52.4 & 40.7 & 37.4 \\ \bottomrule
\end{tabular}
\end{table}

The experimental results indicate that, on the one hand, simply cutting out patches and inferring them separately, as~ELF, is impractical. 
This approach generates a significant number of patches of different sizes, leading to higher function invocation costs. 
On the other hand, simply masking the non-RoIs is also futile because the large resolution slows down the speed of function inference.
Tangram efficiently reduces bandwidth consumption by aligning RoIs into patches, and it lowers function costs by stitching these patches onto a unified canvas, thus accelerating the inference process.

\begin{figure}[!t]
    \centering
    \begin{subfigure}{0.24\textwidth}
        \centering
        \includegraphics[width=1\linewidth]{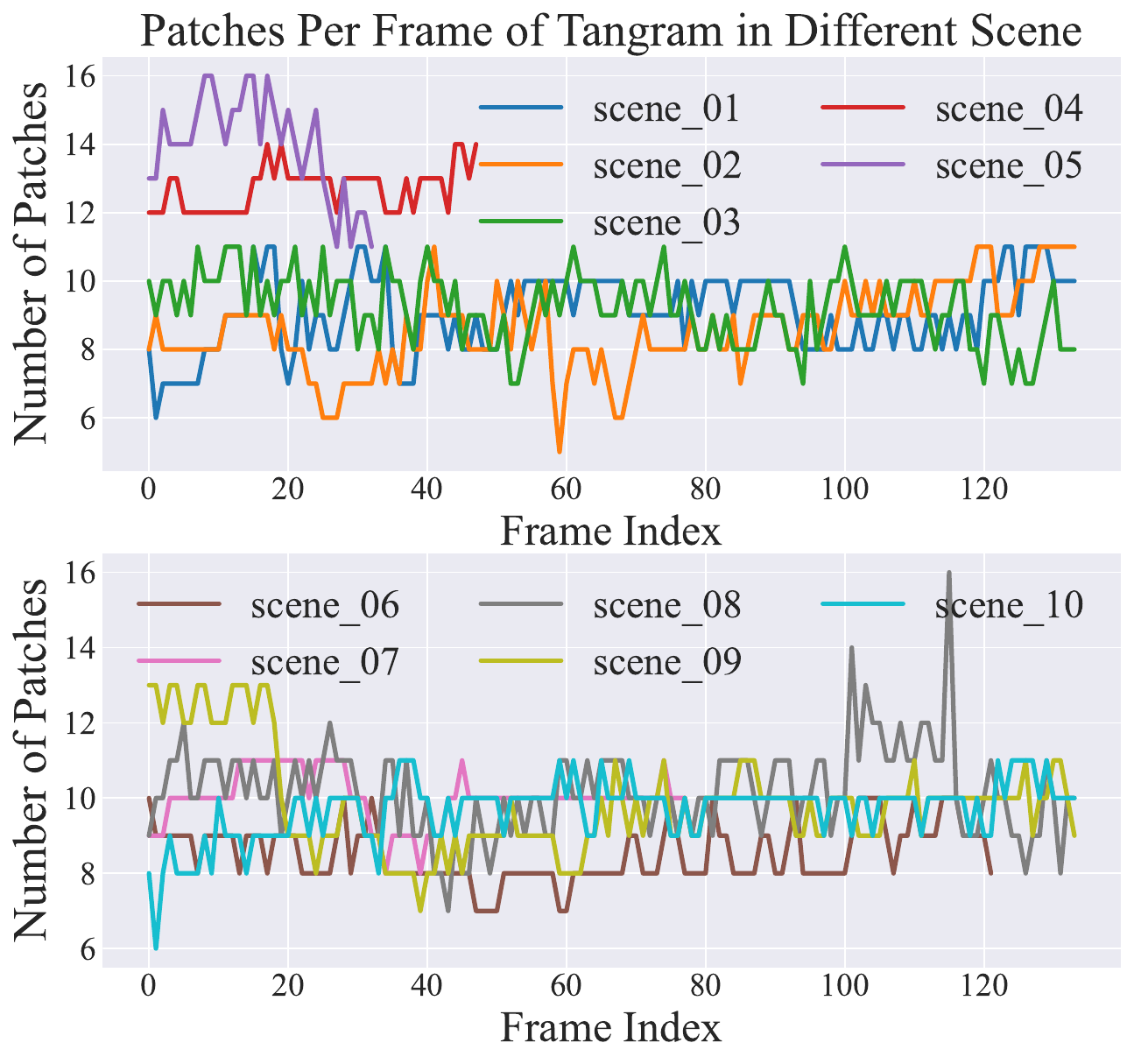}
        \caption{Patches Per Frame}
        \label{fig:sub1}
    \end{subfigure}
        \hfill
      \begin{subfigure}{0.24\textwidth}
        \centering
        \includegraphics[width=1\linewidth]{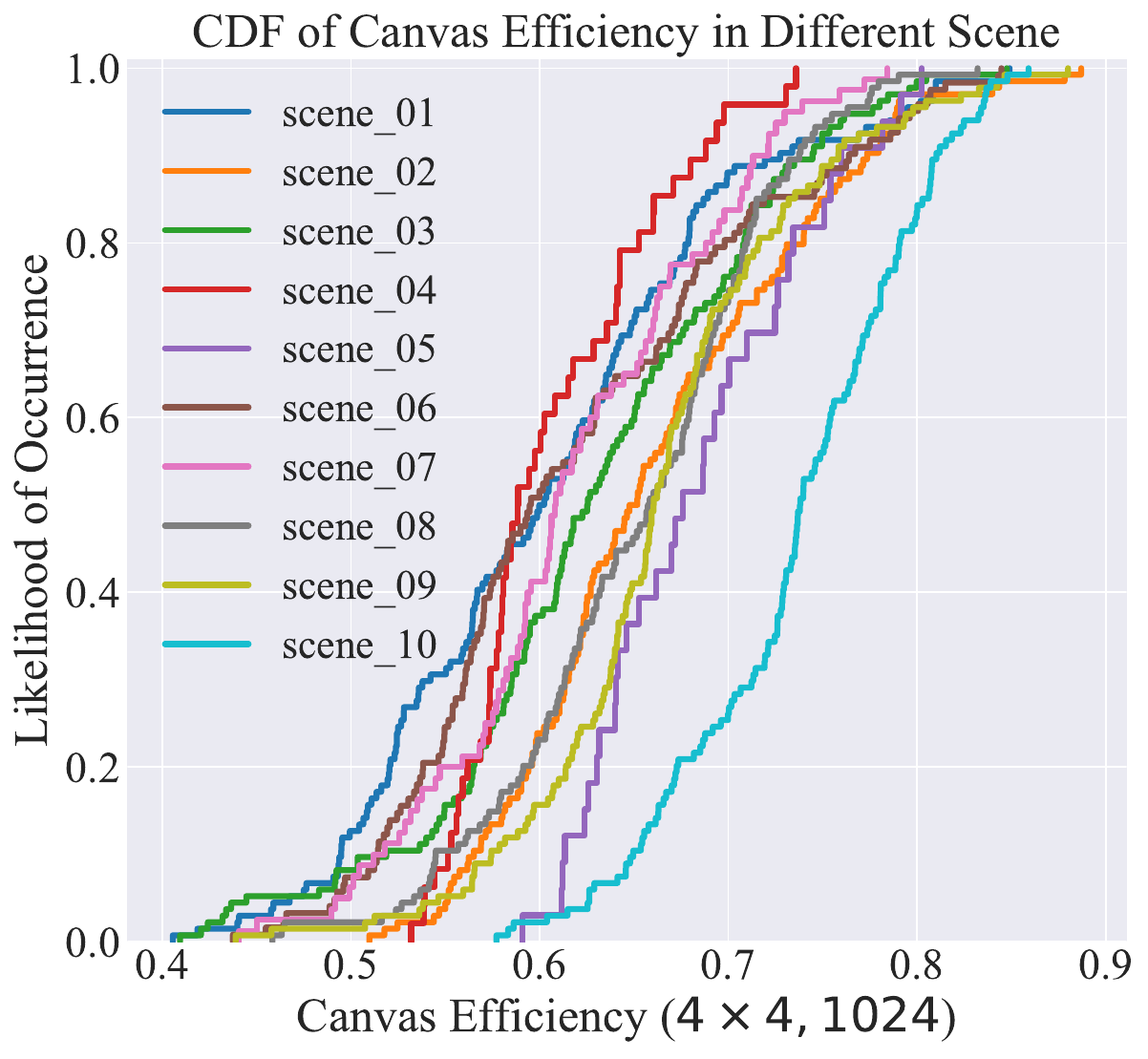}
        \caption{CDF of Canvas Efficiency}
        \label{fig:sub2}
    \end{subfigure}
    \caption{We implement $4\times4$ adaptive frame partitioning algorithm on PANDA4k dataset. (a) shows the patch number generated in each frame. (b) depicts the CDF of canvas efficiency.}
    \label{fig7}
\end{figure}

\begin{figure}[htbp]
    \centering
    \begin{subfigure}{0.24\textwidth}
        \centering
        \includegraphics[width=1\linewidth]{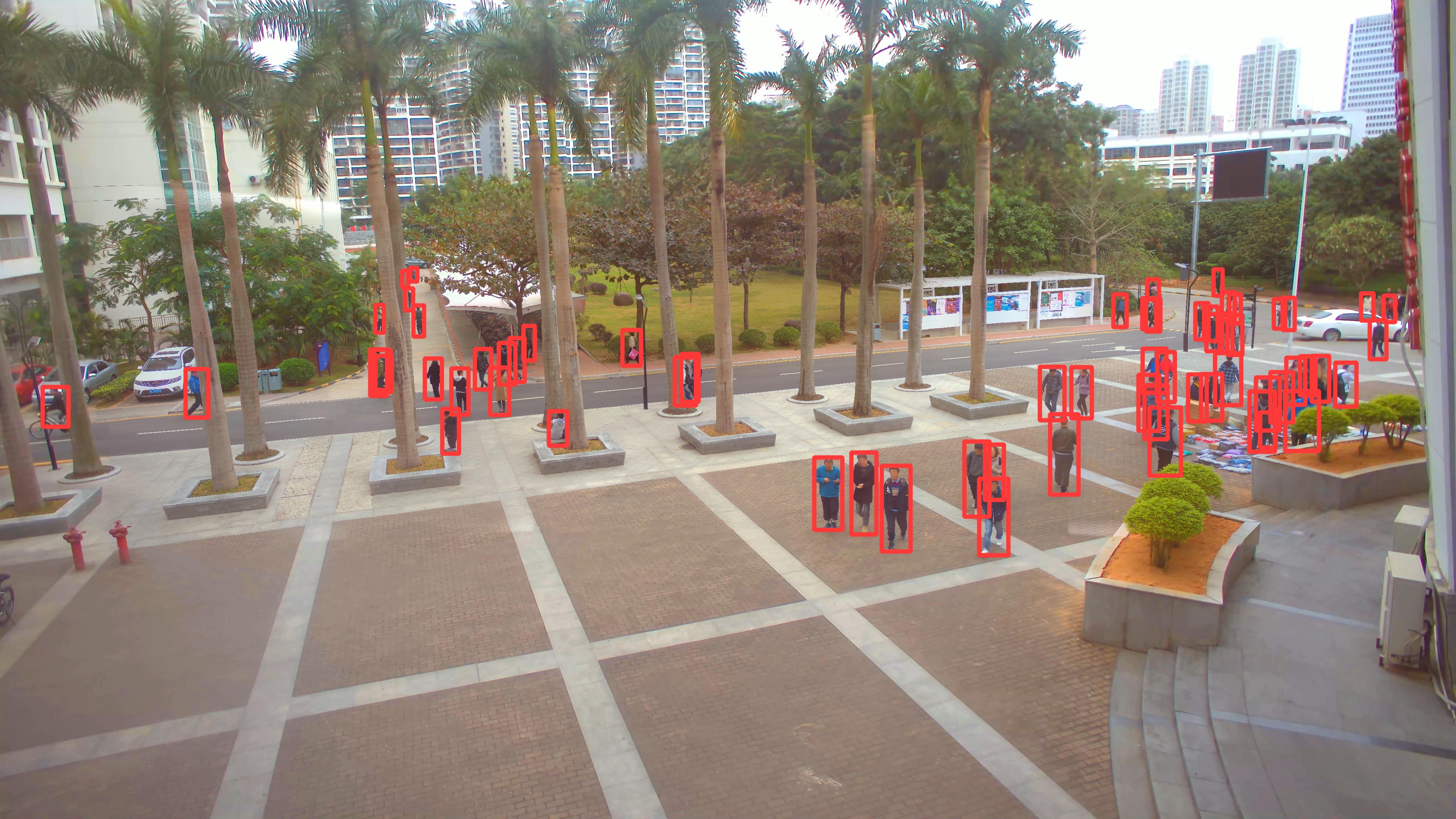}
        \caption{Scene\_01 Frame\#101}
        \label{fig8:sub1}
    \end{subfigure}
        \hfill
      \begin{subfigure}{0.24\textwidth}
        \centering
        \includegraphics[width=1\linewidth]{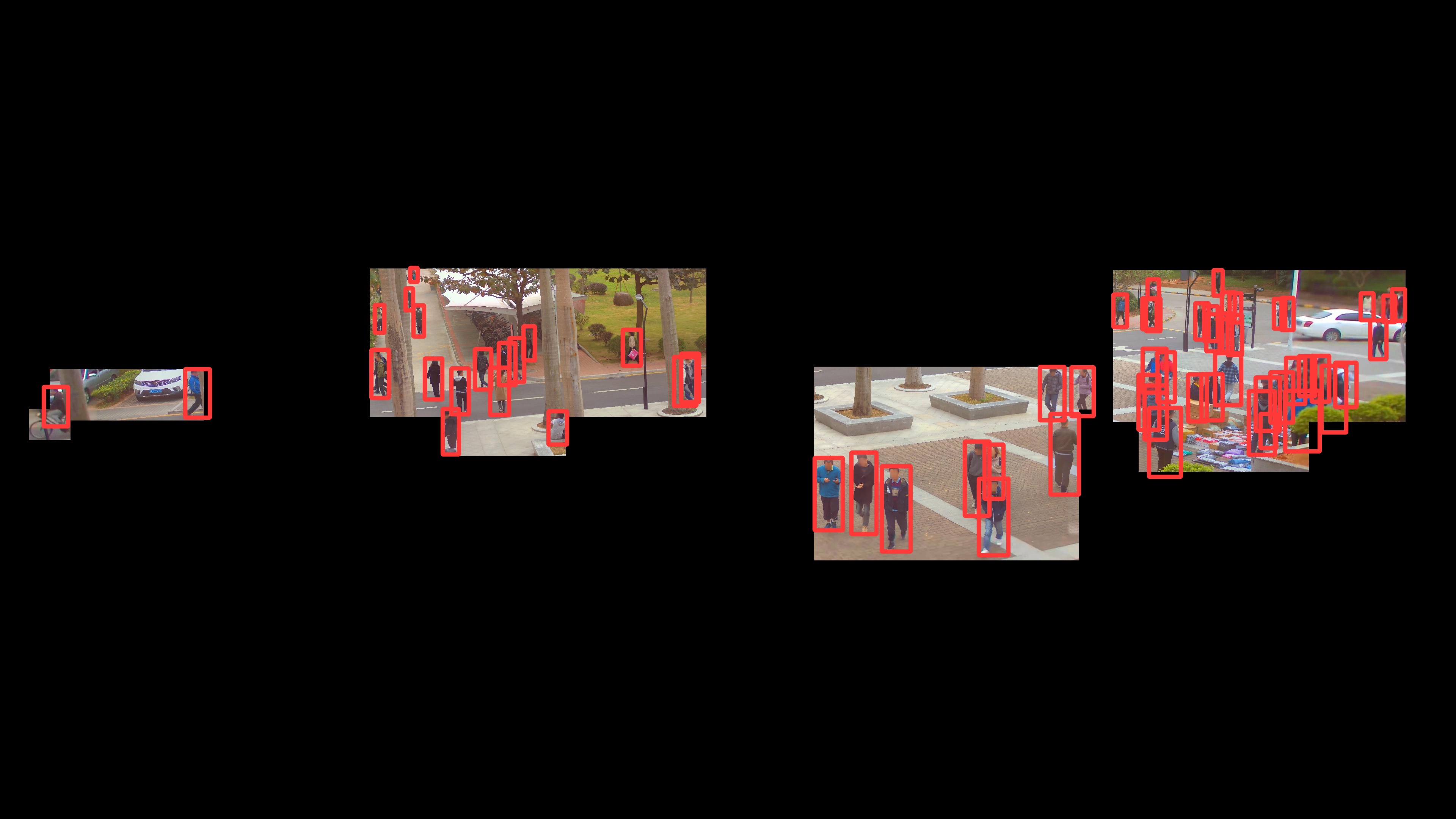}
        \caption{Patches in Frame\#101}
        \label{fig8:sub2}
    \end{subfigure}
       \begin{subfigure}{0.24\textwidth}
        \centering
        \includegraphics[width=1\linewidth]{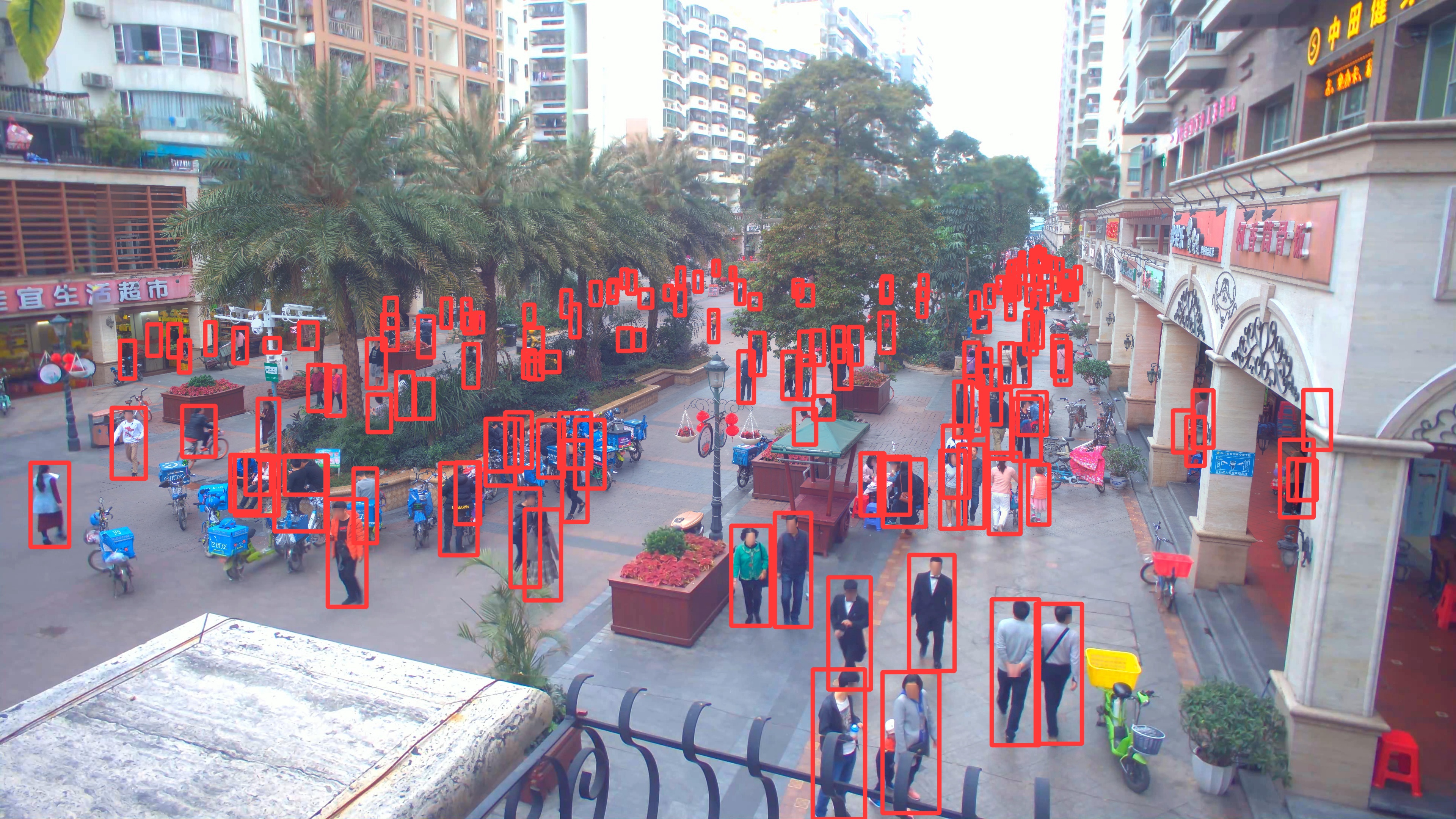}
        \caption{Scene\_08 Frame\#229}
        \label{fig8:sub3}
    \end{subfigure}
        \hfill
      \begin{subfigure}{0.24\textwidth}
        \centering
        \includegraphics[width=1\linewidth]{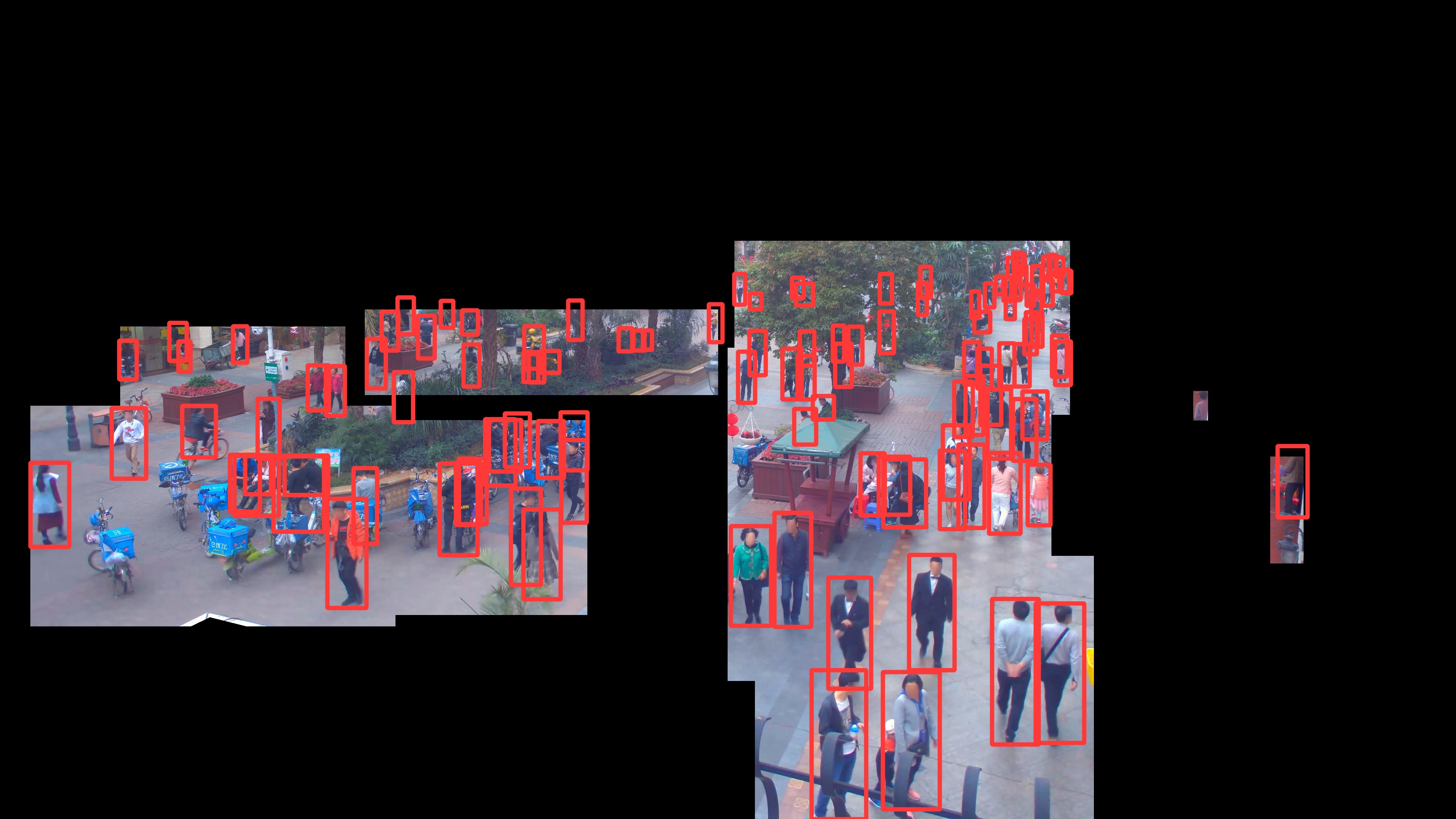}
        \caption{Patches in Frame\#229}
        \label{fig8:sub4}
    \end{subfigure}
    \caption{Example of adaptive frame partitioning algorithm. The red boxes in the figure represent the object. (a)-(d) represents the inference results of a trained Yolov8x model. }
    \label{fig8}
    \vspace{-3.5mm}
\end{figure}

Fig.~\ref{fig7} demonstrates how our algorithm adapts to the dynamic characteristics of inference workload. 
Fig.~\ref{fig7}(a) illustrates the number of patches cut from each frame across ten different scenes, which correlates with the number of objects and their density.
For example, in scene\_01, the 101st frame (see Fig.~\ref{fig8}(a) and \ref{fig8}(b)), the algorithm only needs to generate eight patches due to the relatively small number and intensive distribution of objects. 
However, in the 229th frame of scene\_08 (see Fig.~\ref{fig8}(c) and \ref{fig8}(d)), the objects are distributed across most regions of the frame. 
Therefore, a larger number (i.e., 11) of patches are generated to contain them. 
Using the adaptive frame partitioning algorithm, Tangram can adapt to the changing number and positions of objects, thereby partitioning the most suitable patches and reducing unnecessary bandwidth consumption.

\begin{figure}[!t]
\begin{center}
\includegraphics[width=1\linewidth]{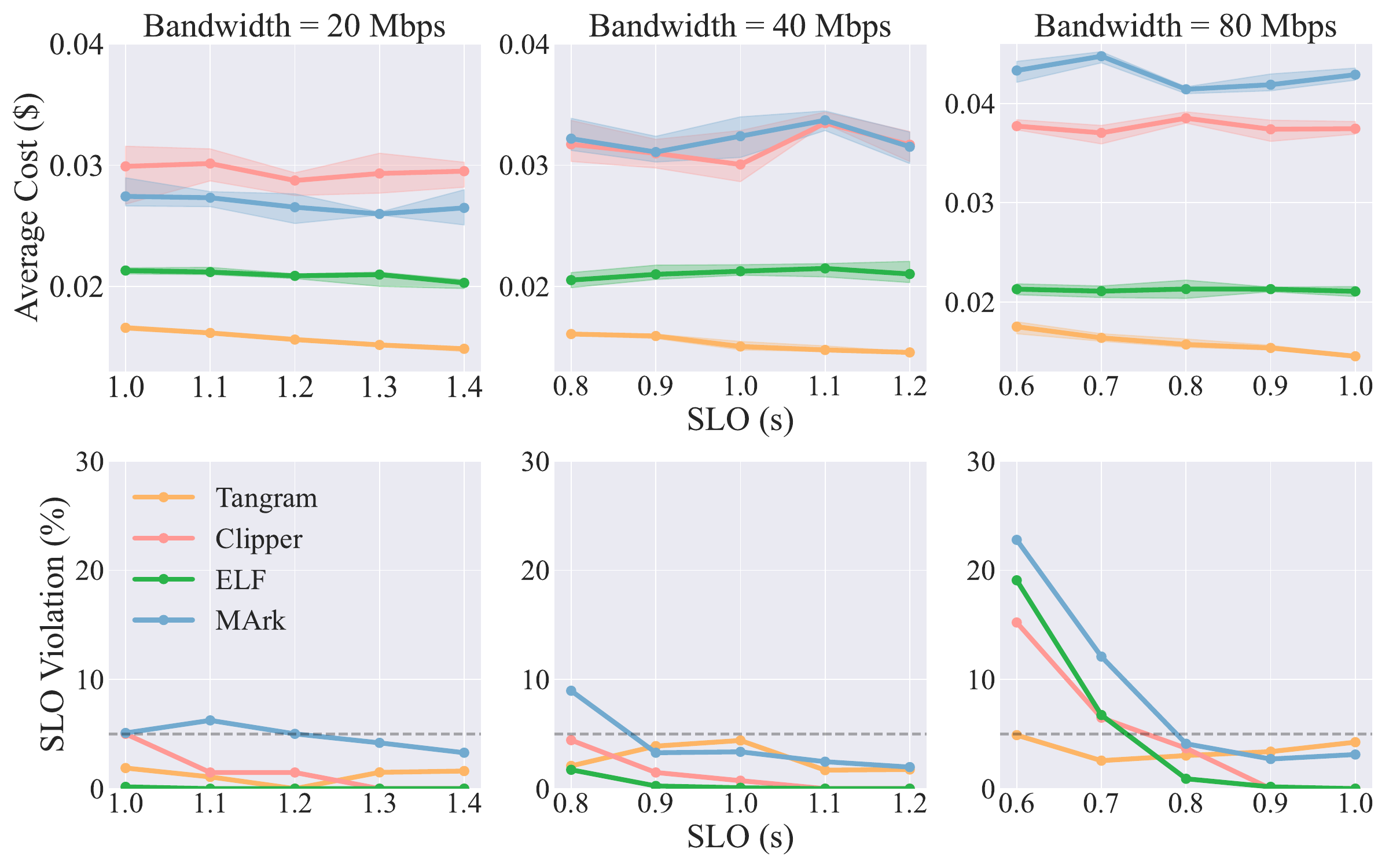}
\caption{The end-to-end performance of Tangram.}
\label{fig:mainexperiment}
\end{center}
\end{figure}

Next, we show the end-to-end performance of Tangram. We set the bandwidth to 20Mbps, 40Mbps, and 80Mbps to simulate different arrival speeds of patches.
We evaluate the cost and SLO violation of Tangram under different SLO restrictions.
Under each bandwidth and SLO configuration shown in Fig.~\ref{fig:mainexperiment}, Tangram achieves the lowest cost and keeps the violation rate below 5\%. 
Specifically, Tangram saves costs up to 61.20\%, 31.03\%, and 66.35\% compared to Clipper, ELF, and MArk under three bandwidth configurations, respectively. 
With the careful design of the scheduler, users no longer need to care about current bandwidth. 
They only need to provide an SLO, and Tangram will adjust the batch size to minimize costs. 
The applications that are highly sensitive to the SLO can manually adjust the slack time $T_{slack}$ to a more conservative estimation.

\begin{table*}[htbp]
\caption{Comparisons of Inference Accuracy (AP)}
\label{table3}
\begin{tabular}{@{}c|cccc|c|cccc@{}}
\toprule
\multirow{2}{*}{\textbf{Scene}} & \multicolumn{4}{c|}{\textbf{Accuracy (AP)}} & \multirow{2}{*}{\textbf{Scene}} & \multicolumn{4}{c}{\textbf{Accuracy (AP)}} \\ \cmidrule(lr){2-5} \cmidrule(l){7-10} 
 & \multicolumn{1}{c|}{\textbf{Full}} & \multicolumn{1}{c|}{\textbf{Partitions (2x2)}} & \multicolumn{1}{l|}{\textbf{Partitions (4x4)}} & \multicolumn{1}{l|}{\textbf{Partitions (6x6)}} &  & \multicolumn{1}{c|}{\textbf{Full}} & \multicolumn{1}{c|}{\textbf{Partitions (2x2)}} & \multicolumn{1}{l|}{\textbf{Partitions (4x4)}} & \multicolumn{1}{l}{\textbf{Partitions (6x6)}} \\ \midrule
\textbf{01} & \multicolumn{1}{c|}{0.572} & \multicolumn{1}{c|}{\textbf{0.583 (+0.011)}} & \multicolumn{1}{c|}{0.573 (+0.001)} & 0.565 (-0.007) & \textbf{06} & \multicolumn{1}{c|}{0.686} & \multicolumn{1}{c|}{\textbf{0.665 (-0.021)}} & \multicolumn{1}{c|}{0.647 (-0.039)} & 0.644 (-0.042) \\
\textbf{02} & \multicolumn{1}{c|}{0.767} & \multicolumn{1}{c|}{\textbf{0.756 (-0.011)}} & \multicolumn{1}{c|}{0.747 (-0.020)} & 0.750 (-0.017) & \textbf{07} & \multicolumn{1}{c|}{0.698} & \multicolumn{1}{c|}{0.663 (-0.035)} & \multicolumn{1}{c|}{\textbf{0.692 (-0.006)}} & 0.672 (-0.026) \\
\textbf{03} & \multicolumn{1}{c|}{0.576} & \multicolumn{1}{c|}{\textbf{0.570 (-0.006)}} & \multicolumn{1}{c|}{0.549 (-0.027)} & 0.493 (-0.083) & \textbf{08} & \multicolumn{1}{c|}{0.638} & \multicolumn{1}{c|}{\textbf{0.626 (-0.012)}} & \multicolumn{1}{c|}{0.622 (-0.016)} & 0.549(-0.089) \\
\textbf{04} & \multicolumn{1}{c|}{0.964} & \multicolumn{1}{c|}{0.962 (-0.002)} & \multicolumn{1}{c|}{\textbf{0.964 (0)}} & 0.927 (-0.037) & \textbf{09} & \multicolumn{1}{c|}{0.598} & \multicolumn{1}{c|}{0.587 (-0.011)} & \multicolumn{1}{c|}{\textbf{0.598 (0)}} & 0.553 (-0.045) \\
\textbf{05} & \multicolumn{1}{c|}{0.899} & \multicolumn{1}{c|}{0.893 (-0.006)} & \multicolumn{1}{c|}{\textbf{0.894 (-0.005)}} & 0.830 (-0.069) & \textbf{10} & \multicolumn{1}{c|}{0.634} & \multicolumn{1}{c|}{\textbf{0.615 (-0.019)}} & \multicolumn{1}{c|}{\textbf{0.615 (-0.019)}} & 0.586 (-0.048) \\ \bottomrule
\end{tabular}
\end{table*}

\subsection{Deep Dive into the Tangram}

\begin{figure}[!t]
\begin{center}
\includegraphics[width=1\linewidth]{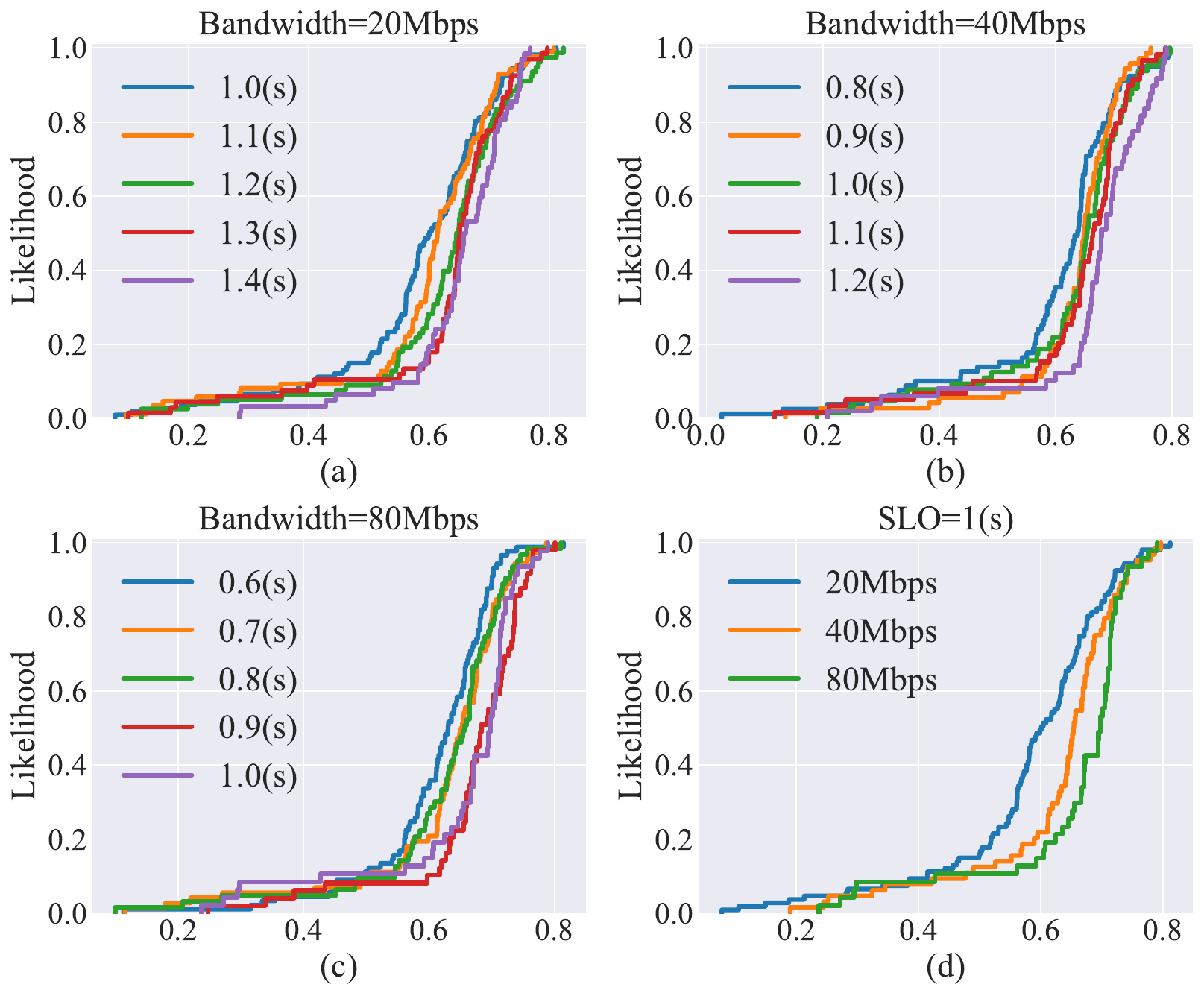}
\caption{Canvas efficiency is influenced by the configurations of bandwidth and SLOs}
\label{fig:why}
\end{center}
\end{figure}

In this subsection, we analyze the Tangram thoroughly and reveal some interesting insights. 
Fig.~\ref{fig:why} shows the CDF of canvas efficiency (i.e., the ratio of the total patch areas to the canvas area) under different configurations of bandwidth and SLOs. 
The reason why the cost of Tangram in Fig.~\ref{fig:mainexperiment} exhibits a decreasing trend as the SLO becomes larger is that the canvas's efficiency is increasing.
Specifically, Fig.~\ref{fig7}(b) and Fig.~\ref{fig:why}(a-c) support this conclusion by showing that as the SLO increases, the average canvas efficiency of each batch also increases because Tangram has more time to wait for the next patch to stitch them into the unfilled canvas, leading to a higher GPU utilization. 
Fig.~\ref{fig:why}(d) confirms this point of view from the bandwidth perspective. 
Under the same SLO constraint, a higher bandwidth implies a higher rate of patch arrival, providing the stitching algorithm with more choices. 
For example, in the case of 20Mbps bandwidth, only 50\% of the canvas efficiency is over 60\%. But when the bandwidth increases to 40Mbps and 80Mbps, approximately 80\% and 86\% of the canvas efficiency are above 60\%, respectively.

\begin{figure}[!t]
\begin{center}
\includegraphics[width=1\linewidth]{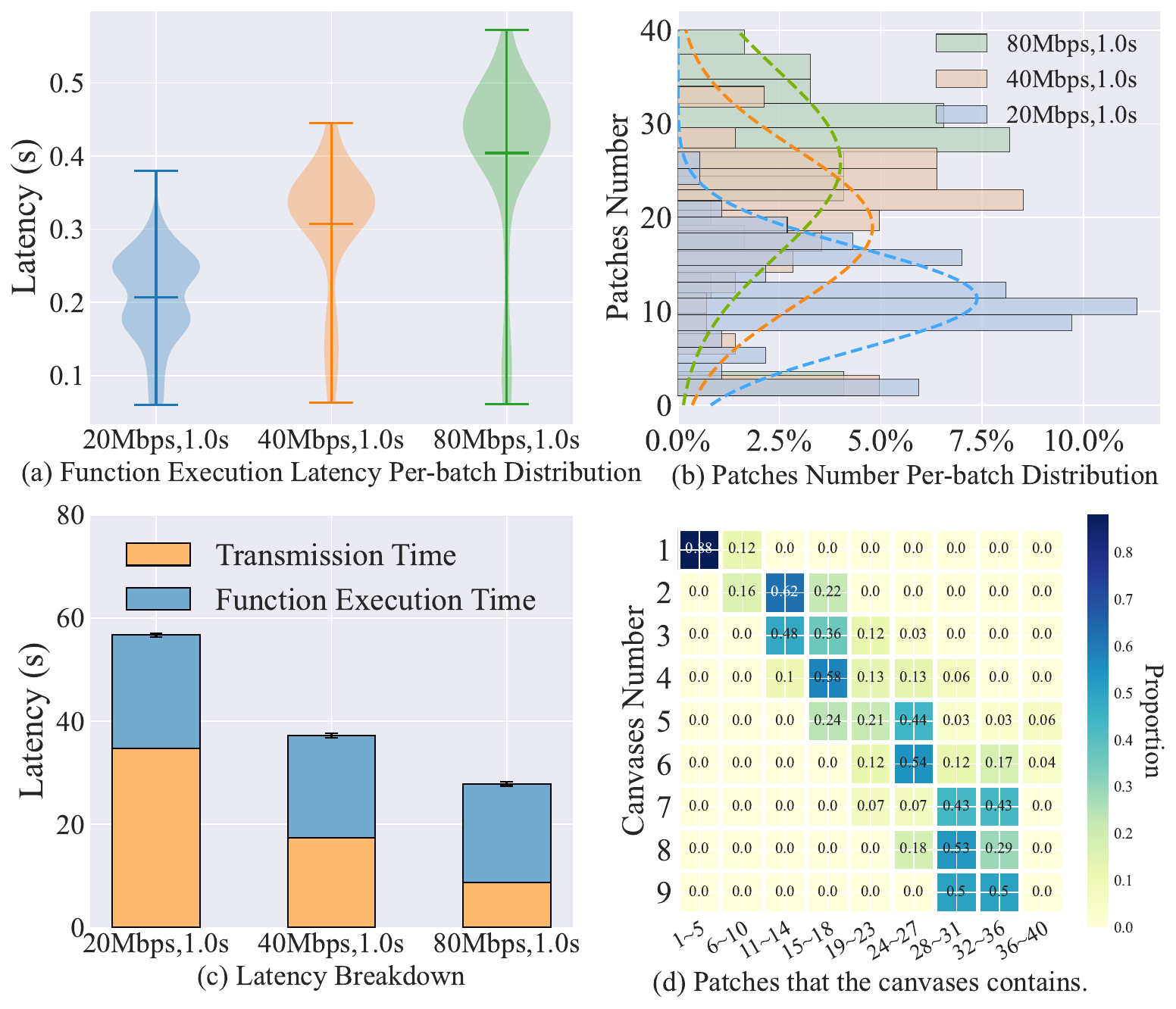}
\caption{Illustration of Tangram insight.}
\label{fig:info}
\vspace{-2mm}
\end{center}
\end{figure}

Another critical observation is that Tangram can maximize the number of patches stitched into a single batch as much as possible as long as it satisfies the SLO, thereby amortizing the cost and latency of each patch. 
We set the SLO as~$1.0$s.
Fig.~\ref{fig:info}(a) displays the distribution of function execution latency for each batch request under three different bandwidth configurations, and Fig.~\ref{fig:info}(c) shows the latency breakdown, including the total transmission time and the total function execution time.
Fig.~\ref{fig:info}(b) illustrates the distribution of patch quantities in each batch. 
As a result, the three subfigures show that although execution time per batch is larger with higher bandwidth, the amortized average latency per patch is reduced. 
Specifically, under the three bandwidth configurations, the amortized average latency per patch is calculated as 0.0252s, 0.0223s, and 0.0213s, respectively.
Finally, Fig.~\ref{fig:info}(d) the probability distribution for the number of patches (represented on the x-axis) contained by varying numbers of canvases (represented on the y-axis) in each batch at 80Mbps bandwidth.
The number of patches and canvases exhibits a positive correlation.
%

\subsection{Accuracy}\label{accuracy}
\begin{table}[]
\centering
\begin{tabular}{@{}c|c|c|c@{}}
\toprule
                               \textbf{Method}   & \textbf{RoI} & \textbf{+Partition}  & \textbf{BW Cons.} \\ \midrule
\multicolumn{1}{c|}{\textbf{GMM~\cite{stauffer1999adaptive}}} & 0.515        & 0.678                  & 67.99\%                        \\
\textbf{Optical Flow~\cite{farneback2003two}}          &0.480         & 0.669                  &77.27\%                        \\
\textbf{SSDLite-MobileNetV2~\cite{Liu_2016}}          & 0.436        & 0.637                 & 82.26\%                        \\
\textbf{Yolov3-MobileNetV2~\cite{redmon2018yolov3}}          & 0.397        & 0.583                & 54.81\%                        \\
\bottomrule
\end{tabular}
\caption{Performance of different RoI extraction methods}
\label{tab:ablation}
\end{table}

After discussing the end-to-end performance of Tangram, it is essential to show that it has negligible impacts on the accuracy of the task. 
The object detection accuracy is mainly affected by the RoI extraction quality. 
An aggressive partition method will make the original frame lose too much information. 
Fortunately, our approach only has a limited impact on accuracy, and the partitioning parameters (i.e., $X, Y$) can be used as a knob to trade off the accuracy and the bandwidth consumption. 
Table \ref{table3} reports the accuracy in different partitioning settings. 
We use average precision ($AP_{.50}$) as the metric. 
A higher average precision indicates better precision and recall performance for the object detection algorithm. 
Our method exhibits accuracy losses of no more than 4\%, 5\%, and 9\%, respectively, when configured with parameters of $2\times2$, $4\times4$, and $6\times6$.
This is because the finer the division of zones, the greater the likelihood of potential objects being lost between zones.

Last, we compare the performance of several RoI extraction methods. 
Gunnar Farneback's algorithm~\cite{farneback2003two} computes the optical flow for each pixel, conveniently enabling the extraction of moving RoIs between two consecutive frames.
SSDLite-MobileNetV2~\cite{Liu_2016} and Yolov3-MobileNetV2~\cite{redmon2018yolov3} are two learning-based lightweight vision models, and we use their pre-trained models for RoI extraction.
Table~\ref{tab:ablation} presents the accuracy by only applying different RoI detection methods, the accuracy by applying our adaptive frame partitioning algorithm in different RoI detection methods  (Partition), and the proportion of bandwidth consumption (BW Cons.).
Note that a full frame detection has an AP of 0.60 in the experiment. In this work, we select GMM because of its effective trade-off between accuracy and bandwidth consumption.
\section{Related Work}\label{sec_VII}

In this section, we start with a brief review of video analysis systems in cloud-edge environments, followed by an in-depth exploration of literature about serverless architectures. Finally, we delve into the application of batching within serverless platforms.

\subsection{Video Analytics System}
DAO~\cite{murad2022dao} is a dynamic adaptive offloading framework for video analytics. It dynamically adjusts the bitrate and resolution of video offloading to enhanced inference precision. 
To mitigate inference latency and reduce energy or bandwidth overhead, JAVP~\cite{yang2023javp} and DCSB~\cite{cao2023edge} determine the inference routing based on the difficulty level of the video input. 
SmartFilter~\cite{tchaye2022smartfilter}, guided by a reinforcement learning model, identifies keyframes in the video and offloads them to the cloud for model inference for better efficiency.
Similar researchs~\cite{wang2019bridging,elf,vabus,chen2022context,eagleeye} also aim to reduce the computation of video processing while maintaining high accuracy.

Thanks to the rapid and elastic scalability and a pay-as-you-go billing model of the serverless, many video analytics systems have been migrated to the cloud now~\cite{zhang2021serverless}. 
CEVAS~\cite{zhang2021towards} is a cloud-edge video analytics system that leverages the serverless computing paradigm to tackle the online video query pipelines. 
It predicts resource usage based on video characteristics and partitions the pipeline with a directed acyclic graph structure between the edge and the cloud. 
VPaaS~\cite{zhangserverless} is a serverless cloud-fog platform that minimizes the cloud infrastructure cost and bandwidth usage while maintaining high accuracy in various video applications. 
LLAMA~\cite{romero2021llama} is a serverless framework that accommodates heterogeneous hardware and automatically optimizes each operation knob and resource allocation options to achieve various latency targets through 5 typical video analytics pipelines. 
Literature~\cite{wang2023edge} studies the problem of optimal dynamic configuration in serverless-based video analytics systems.
However, Tangram is dedicated to bandwidth optimization and cost reduction in high-resolution video applications in serverless platforms.

\subsection{Batching in Serverless Platform}
Batching is an essential operation for ML model serving and serverless functions. Clipper~\cite{crankshaw2017clipper} and MArk~\cite{Zhang2020EnablingCS} introduce the batch size and timeout parameters to control the batching. 
BATCH~\cite{ali2020batch} establishes a Markov-modulated Poisson Process to capture the request arrival process and optimize the configuration parameters (i.e., memory size, batch size, and timeout) to minimize the cost while satisfying SLO. 
MBS~\cite{ali2022optimizing} is a similar framework for serving heterogeneous ML inference workloads with SLO guarantees for NLP applications. 
OTAS~\cite{chen2024otas} groups queries with similar arrival patterns and SLOs into batches, then allocating different inference configurations to each batch.
However, unlike the aforementioned batching strategies, Tangram ingeniously integrates the inference batch into the stitching operation without manipulating the batch size and timeout parameters.

\section{Conclusion}\label{sec_VIII}
We design Tangram, a video analytics system that takes advantage of several techniques to optimize the cost of high-resolution video analytics in the cloud-edge scenario. 
This system minimizes the cost of DNN inference based on serverless functions while satisfying SLO requirements. 
The main contribution stems from the novel approach of stitching-based batch processing and the online SLO-aware batching algorithm. 
Our study shows that Tangram can reduce bandwidth consumption and cost up to 74.30\% and 66.35\% while maintaining SLO violations within 5\% and the accuracy loss negligible.

\newpage
\bibliographystyle{IEEEtran}
\bibliography{infocom}

\end{document}